\documentclass{LMCS}

\usepackage{amsmath,amssymb,amsthm,mylogic,hyperref}

\newcommand{\NN}{{\mathbb N}}
\newcommand{\RR}{{\mathbb R}}
\newcommand{\QQ}{{\mathbb Q}}
\newcommand{\CC}{{\mathbb C}}
\newcommand{\ZZ}{{\mathbb Z}}
\newcommand{\AAA}{{\mathbb A}}

\newcommand{\Tadd}{T_{\fn{add}}}
\newcommand{\Tmult}{T_{\fn{mult}}}
\newcommand{\Tcommon}{T_\fn{comm}}

\newcommand{\phadd}{\ph_{\fn{add}}}
\newcommand{\phmult}{\ph_{\fn{mult}}}

\newcommand{\Deltaadd}{\Delta_{\fn{add}}}
\newcommand{\Deltamult}{\Delta_{\fn{mult}}}
\newcommand{\Deltacommon}{\Delta_\fn{comm}}

\newcommand{\RRR}{\mdl R}
\newcommand{\dom}{\fn{dom}}
\newcommand{\rng}{\fn{rng}}
\newcommand{\fld}{\fn{fld}}
\newcommand{\alg}{\fn{alg}}
\newcommand{\trans}{\fn{trans}}

\renewcommand{\ln}{\fn{ln}}
\renewcommand{\exp}{\fn{exp}}

\newcommand{\timesstar}{\otimes}

\def\doi{2 (4:4) 2006}
\lmcsheading%
{\doi}
{1--42}
{}
{}
{Jan.~31, 2006}
{Oct.~18, 2006}
{}

\begin{document}

\title{Combining decision procedures for the reals}

\author[J.~Avigad]{Jeremy Avigad\rsuper{a}}
\address{{\lsuper a}Department of Philosophy, Carnegie Mellon
University, Pittsburgh, PA 15213}
\email{avigad@cmu.edu}

\author[H.~Friedman]{Harvey Friedman\rsuper{b}}
\address{{\lsuper b}Department of Mathematics, Ohio State University, Columbus, OH 43210}
\email{friedman@math.ohio-state.edu}

\thanks{{\lsuper{a,b}}Work by both authors has been supported by NSF grant DMS-0401042. We
are grateful to three anonymous referees for numerous comments,
suggestions, and corrections.}


\begin{abstract}
  We address the general problem of determining the validity of
  boolean combinations of equalities and inequalities between
  real-valued expressions. In particular, we consider methods of
  establishing such assertions using only restricted forms of
  distributivity. At the same time, we explore ways in which ``local''
  decision or heuristic procedures for fragments of the theory of the
  reals can be amalgamated into global ones.

  Let $\Tadd[\QQ]$ be the first-order theory of the real numbers in
  the language with symbols $0, 1, +, -, <, \ldots, f_a, \ldots$ where
  for each $a \in \QQ$, $f_a$ denotes the function $f_a(x) = ax$.  Let
  $\Tmult[\QQ]$ be the analogous theory for the language with symbols
  $0, 1, \times, \div, <, \ldots, f_a, \ldots$.  We show that although
  $T[\QQ] = \Tadd[\QQ] \cup \Tmult[\QQ]$ is undecidable, the universal
  fragment of $T[\QQ]$ is decidable. We also show that terms of $T[\QQ]$
  can fruitfully be put in a normal form. We prove analogous results for
  theories in which $\QQ$ is replaced, more generally, by suitable
  subfields $F$ of the reals. Finally, we consider practical
  methods of establishing quantifier-free validities that
  approximate our (impractical) decidability results.
\end{abstract}

\maketitle


\section{Introduction}
\label{introduction:section}

This paper is generally concerned with the problem of determining
the validity of boolean combinations of equalities and inequalities
between real-valued expressions. Such computational support is
important not only for the formal verification of mathematical proofs,
but, more generally, for any application which depends on such
reasoning about the real numbers.

Alfred Tarski's proof \cite{tarski:51} that the theory of the real
numbers as an ordered field admits quantifier-elimination is a
striking and powerful response to the problem. The result implies
decidability of the full first-order theory, not just the
quantifier-free fragment. George Collins's \cite{collins:75} method of
cylindrical algebraic decomposition made this procedure feasible in
practice, and ongoing research in computational real geometry has
resulted in various optimizations and alternatives (see
e.g.~\cite{dolzmann:et:al:98,basu:et:al:03,basu:99}). Recently, a
proof-producing version of an elimination procedure due to Paul Cohen
has even been implemented in the framework of a theorem prover for
higher-order logic \cite{mclaughlin:harrison:05}.

There are two reasons, however, that one might be interested in
alternatives to q.e.\ procedures for real closed fields. The first is
that their generality means that they can be inefficient in restricted
settings. For example, one might encounter an inference like
\[
0 < x < y \rightarrow (1 + x^2) / (2 + y)^{17} < (1 + y^2) / (2 +
x)^{10},
\]
in an ordinary mathematical proof. Such an inference is easily
verified, by noticing that all the subterms are positive and then
chaining through the obvious inferences. Computing sequences of
partial derivatives, which is necessary for the full decision
procedure, seems misguided in this instance. A second, more compelling
reason to explore alternatives is that decision procedures for real
closed fields are not extensible. For example, adding trigonometric
functions or an uninterpreted unary function symbol renders the full
first-theory undecidable. Nonetheless, an inference like
\[
0 < x < y \rightarrow (1 + x^2) / (2 + e^y) < (2 + y^2) / (1 + e^x)
\]
is also straightforward, and it is reasonable to seek
procedures that capture such inferences.

The unfortunate state of affairs is that provability in most
interesting mathematical contexts is undecidable, and even when
decision procedures are available in restricted settings, they are
often infeasible or impractical. This suggests, instead, focusing on
heuristic procedures that traverse the search space by applying a
battery of natural inferences in a systematic way (for some examples
in the case of real arithmetic, see
\cite{beeson:98,hunt:et:al:03,tiwari:03}). There has been,
nonetheless, a resistance to the use of such procedures in the
automated reasoning community. For one thing, they do not come with a
clean theoretical characterization of the algorithm's behavior, or the
class of problems on which one is guaranteed success.  This is closely
linked to the fact that the algorithms based on heuristics are
brittle: small changes and additions as the system evolves can have
unpredictable effects.

The strategy we pursue here is to develop a theoretical understanding
that can support the design of such heuristic procedures, by
clarifying the possibilities and limitations that are inherent in a
method, and providing a general framework within which to situate
heuristic approaches. One observation we exploit here is that
often distributivity is used only in restricted ways in the types of
verifications described above. Arguably, any inference that
requires factoring a complex expression does not count as ``obvious.''
Conversely, multiplying through a sum can result in the loss of
valuable information, as well as lead to increases in the lengths of
terms. As a result, steps like these are usually spelled out
explicitly in textbook reasoning when they are needed. It is therefore
natural to ask whether one can design procedures that reasonably
handle those inferences that do not make use of distributivity,
relying on the user or other methods to then handle the latter.

The ``distributivity-free'' fragment of the theory of the reals as an
ordered field can naturally be viewed as a combination of the additive
and multiplicative fragments, each of which is easily seen to be
decidable. This points to another motivation for our approach. A
powerful paradigm for designing useful search procedures involves
starting with procedures that work locally, for restricted theories,
and then amalgamating them into a global procedure in some principled
way. For example, Nelson-Oppen methods are currently used to combine
decision procedures for theories that are disjoint except for the
equality symbol, yielding decision procedures for the universal
fragment of their union. Shostak methods perform a similar task more
efficiently by placing additional requirements on the theories to be
amalgamated. (See \cite{janacic:bundy:02,ranise:et:al:04} and the
introduction to \cite{baader:et:al:04} for overviews of the various
approaches.) Such methods are appealing, in that they allow one to
unify such decision procedures in a uniform and modular way.  This
comes closer to what ordinary mathematicians do: in simple,
domain-specific situations, we know exactly how to proceed, whereas in
more complex situations, we pick out the fragments of a problem that we
know how to cope with and then try to piece them together. One would
therefore expect the notion of amalgamating decision procedures,
or even heuristic procedures, to be useful when there is more
significant overlap between the theories to be amalgamated.  For
example, the Nelson-Oppen procedure has been generalized in various
ways, such as to theories whose overlap is ``locally finite''
\cite{ghilardi:04}.  Our results here show what can happen when one
tries to amalgamate decision procedures for theories where the
situation is not so simple.

Sections~\ref{combining:section} and \ref{fragments:section}, below,
provide general background. In Section~\ref{combining:section}, we
discuss the theoretical results that underly Nelson-Oppen methods for
combining decision procedures for theories that share only the
equality symbol, or for theories with otherwise restricted overlap. In
Section~\ref{fragments:section}, we describe some particular decision
procedures for fragments of the reals, which are candidates for such a
combination.

In Section~\ref{theories:section}, we define the theories $T[F]$,
which combine the additive and multiplicative fragments of the theory of
the reals, allowing multiplicative constants from a field $F$. The
theory $T[F]$, in particular, can, alternatively, be thought of as the
theory of real closed fields minus distributivity, except for
constants in $F$. Because of the nontrivial overlap, Nelson-Oppen
methods no longer apply. In Section~\ref{examples:section}, we provide
two examples that clarify what these theories can do. On the positive
side, we show that when a multivariate polynomial has no roots on a
compact cube, $T[\QQ]$ is strong enough to prove that fact.  On the
negative side, we show that the theories $T[F]$ cannot prove $x^2 - 2x
+ 1 \geq 0$, a fact which is easily proved using distributivity.

In Sections~\ref{provability:section}--\ref{normal:forms:section} we
establish our decidability results. Using a characterization of the
universal fragment of $T[F]$ developed in
Section~\ref{provability:section}, we show, in
Section~\ref{decidability:section}, that whenever $F$ is an
appropriately computable subfield of $\RR$, the universal fragment of
$T[F]$ is decidable. So, in particular, the universal fragment of
$T[\QQ]$ is decidable. In Section~\ref{normal:forms:section}, we
describe normal forms for terms of $T[F]$, which make it easy to
determine whether two terms are provably equal. We also show that these
provable equalities are independent of the parts of the theory that
have to do with the ordering.

In Sections~\ref{models:section}--\ref{undecidability:two:section}, we
establish our undecidability results. In Section~\ref{models:section},
we present a flexible technique that will allow us to build suitable
models of the theories $T[F]$. In
Section~\ref{undecidability:one:section}, we use this technique to
reduce the problem of determining the truth of an existential sentence
over the field $F$ to that of the provability of a related formula in
$T[F]$. As a result, if Diophantine equations in the rationals are
unsolvable (which is generally believed to be the case), then so is
the set of existential consequences of $T[\QQ]$. In
Section~\ref{undecidability:two:section}, we reduce the problem of
determining the solvability of a Diophantine equation in the integers
to the provability of a related $\forall \forall \forall
\exists^*$-sentence in any $T[F]$. As a result, we have an
unconditional undecidability result for that fragment.

The procedure implicit in our decidability results is not useful in
practice: it works by reducing the question as to whether a universal
sentence in provable in $T[F]$ to the question as to whether a more
complex sentence in provable in the theory of real closed fields, and
then appeals the the decidability of the latter. In
Sections~\ref{avoiding:section}--\ref{extensions:section}, we consider
the problem of designing pragmatic procedures that approximate our
decidability results, are more flexible than decision procedures for
real closed fields, and work reasonably well on ordinary textbook
inferences. In Section~\ref{avoiding:section}, we suggest a
restriction of the theories $T[F]$ which avoids disjunctive case
splits, which are a key source of infeasibility. In
Section~\ref{heuristic:section}, we describe a search procedure that
works along these lines, making use of the normal forms introduced in
Section~\ref{normal:forms:section}.  In
Section~\ref{extensions:section}, we indicate a number of directions
in which one might extend and improve our crude algorithm.

Finally, in Section~\ref{conclusions:section}, we offer some final
thoughts and conclusions.


\section{Combining decision procedures}
\label{combining:section}

In this section, we briefly review the mathematical foundation for the
Nelson-Oppen combination procedure \cite{nelson:oppen:79}. For more
detail, see
\cite{barrett:02,conchon:krstic:unp,harrison:unp,tinelli:harandi:96};
an important program verification system based on these method is
described in \cite{detlefs:et:al:05}. 

Let $\Delta$ be the set of first-order formulas in the language of
equality asserting that the universe is infinite. A theory $T$ is said
to be \emph{stably infinite} if whenever $T \cup \Delta$ proves a
universal sentence $\ph$, then $T$ proves it as well. Equivalently,
$T$ is stably infinite if whenever a quantifier-free formula is
satisfied in any model of $T$, it is satisfied in some infinite model
of $T$. In particular, if $T$ only has infinite models, then $T$ is
stably infinite.

The Nelson-Oppen procedure for combining decidable theories of
equality is based on the following:
\begin{theorem}
\label{nelson:oppen:thm}
Suppose $T_1$ is a theory in a language $L_1$, $T_2$ is a theory in a
language $L_2$, $T_1$ and $T_2$ are stably infinite, and the languages
$L_1$ and $L_2$ are disjoint except for the equality symbol. Suppose
the universal fragments of $T_1$ and $T_2$ are decidable. Then the
universal fragment of $T_1 \cup T_2$ is decidable.\qed
\end{theorem}

The proof of Theorem~\ref{nelson:oppen:thm} is not difficult.  The
question as to whether $T_1 \cup T_2$ proves a universal formula is
equivalent to the question as to whether it proves the quantifier-free
matrix. (One can treat the free variables as new constants, if one
prefers, but here and below we will speak in terms of proving or
refuting sets of formulas with free variables.) Since any
quantifier-free formula can be put in conjunctive normal form, the
problem reduces to that of determining provability of disjunctions of
literals, or, equivalently, that of determining whether $T_1 \cup T_2$
refutes a conjunction of literals.

Let $\Gamma$ be a set of literals. The first step in the procedure is
to introduce new variables to ``separate terms.'' For example, the
universal closure of a formula of the form $\ph(f(s_1,\ldots,s_k))$ is
equivalent to the universal closure of $x = f(s_1,\ldots,s_k)
\limplies \ph(x)$, where $x$ is a new variable. This is, in turn,
equivalent to the universal closure of $y_1 = s_1 \land \ldots \land
y_k = s_k \land x = f(y_1, \ldots, y_k) \limplies \ph(x)$. By
introducing new variables in this way, we can obtain sets of
equalities $\Pi_1$ and $\Pi_2$ in $L_1$ and $L_2$ respectively, and a
set of literals, $\Pi_3$, in which no function symbols occur, such
that $T_1 \cup T_2$ refutes $\Gamma$ if and only if it refutes $\Pi_1
\cup \Pi_2 \cup \Pi_3$. Let $\Gamma_1$ be $\Pi_1$ together with the
literals in $\Pi_3$ that are in $L_1$, and let $\Gamma_2$ be $\Pi_2$
together with the literals in $\Pi_3$ that are in $L_2$. Then each
$\Gamma_i$ is in the language of $T_i$, and $T_1 \cup T_2$ refutes
$\Gamma$ if and only if $T_1 \cup T_2$ refutes $\Gamma_1 \cup
\Gamma_2$.

By the Craig interpolation theorem, $T_1 \cup T_2$ refutes $\Gamma_1
\cup \Gamma_2$ if and only if there is a quantifier-free
interpolant $\theta$ in the common language (i.e.~involving only the
equality symbol and variables common to both $\Gamma_1$ and
$\Gamma_2$) such that
\[
T_1 \cup \Gamma_1 \proves \theta
\]
and
\[
T_2 \cup \Gamma_2 \cup \{ \theta \} \proves \bot.
\]
By the assumption that $T_1$ and $T_2$ are stably infinite, we can
assume without loss of generality that each includes $\Delta$. Since
the theory of equality in an infinite structure has
quantifier-elimination, $\theta$ is equivalent to a
quantifier-free formula. In fact, we can assume without loss of
generality that $\theta$ is in disjunctive normal form. So we are
looking for a sequence $\theta_1, \ldots, \theta_n$ of finite
conjunctions of literals such that for each $i$,
\[
T_1 \cup \Gamma_1 \proves \theta_1 \lor \ldots \lor \theta_n
\]
and 
\[
T_2 \cup \Gamma_2 \cup \{ \theta_i \} \proves \bot
\]
for each $i$.

Each disjunct $\theta_i$ describes relationships between the variables
$\vec x$ of $\Gamma_1 \cup \Gamma_2$, in the language $L_1 \cap L_2$,
which has only the equality symbol. The key point is this: over
$\Delta$, every ``complete type'' (that is, complete, consistent set
of formulas with free variables $\vec x$) is determined by an
exhaustive description of which of the variables are equal to one
another and which are not. Furthermore, there are only finitely many
such descriptions. Without loss of generality, we can assume that each
$\theta_i$ is of this form, because otherwise it can be rewritten as a
disjunction of such. Thus we simply need to use the decision procedure
for $T_2$ to determine all the complete types $\theta_i$ that can be
refuted by $T_2 \cup \Gamma_2$, and then use the decision procedure
for $T_1$ to determine whether $T_1 \cup \Gamma_1$ proves their
disjunction. Equivalently, we can use the decision procedures to
determine all the complete types that are consistent with either side;
$\Gamma$ can be refuted if and only if there is no complete type that
is consistent with both.

This naive procedure is not very efficient. In fact, the Nelson-Oppen
procedure iteratively searches for a disjunction of equalities
derivable from either $T_1 \cup \Gamma_1$ or $T_2 \cup \Gamma_2$, adds
this disjunction to the hypotheses, and then splits across the cases.
It is not hard to show that this variant is complete; one can view it
in terms using both $T_1 \cup \Gamma_1$ and $T_2 \cup \Gamma_2$ to
derive a sequence of increasingly strong disjunctions of
conjunctions of positive literals, until either a contradiction is
reached or no further strengthening can be found. In the latter case,
one can read off a complete type consistent with both $T_1 \cup
\Gamma_1$ and $T_2 \cup \Gamma_2$. The procedure is much more
efficient if either of the theories $T_i$ is \emph{convex}, that is,
whenever $\ph$ is a conjunction of literals and $T_i \cup \ph \proves
x_1 = y_1 \lor \ldots \lor x_k = y_k$ then $T_i \cup \ph \proves x_i =
y_i$ for some $i$.  The linear theory of the reals has this property,
though the multiplicative theory does not.  Shostak's procedure
provides further optimization under the assumptions that terms in the
theory are ``canonizable'' and ``solvable,'' again, features that are
commonly satisfied.

For future use, we record the effects of ``separating terms,'' as
described above. We no longer assume $L_1$ and $L_2$ are disjoint
languages. 

\begin{proposition}
\label{separating:variables:prop}
Let $\ph$ be any universal sentence in the language $L_1 \cup L_2$.
Then $\ph$ is equivalent to a sentence of the form
\[
\fa {\vec x} (\theta_1(\vec x) \land \theta_2(\vec x) \limplies
\theta_3(\vec x)),
\]
where $\theta_1$ is a conjunction of equalities in $L_1$, $\theta_2$
is a conjunction of equalities in $L_2$, and $\theta_3$ is a
quantifier-free formula in $L_1 \cup L_2$ with no function
symbols. As a result, $\ph$ can be written as a conjunction of
formulas of the form
\begin{equation}
\label{separated:eq}
\fa {\vec x} (\ph_1(\vec x) \lor \ph_2(\vec x)),
\end{equation}
where each $\ph_i$ is a quantifier-free formula in $L_i$. If
all the relation symbols in $L_1 \cup L_2$ are common to both $L_1$
and $L_2$, or if the matrix of $\ph$ is equivalent to a disjunction of
literals, one conjunct of the form (\ref{separated:eq}) suffices.\qed
\end{proposition}


\section{Decision procedures for fragments of the reals}
\label{fragments:section}

The method described in Section~\ref{combining:section} requires only
that the universal fragments of the theories $T_1$ and $T_2$ are
decidable, and that for any sequence of variables, there are only
finitely many complete types in the common language, each of which can
be described by a single quantifier-free formula. In particular, we
have the following:

\begin{theorem}
\label{nelson:oppen:extension:thm}
  Let $T_1$ and $T_2$ be theories extending the theory of dense linear
  orders without endpoints, with only $<$ and $=$ in the common
  language. If the universal fragments of $T_1$ and $T_2$ are
  decidable, then the universal fragment of $T_1 \cup T_2$ is also
  decidable.\qed
\end{theorem}

 As was the case when equality was the only common symbol,
this theorem can be stated even more generally: we only need assume
that $T_1$ and $T_2$ satisfy the property obtained by replacing
$\Delta$ by the theory of dense linear orders without endpoints in the
definition of ``stably infinite'' above. Of course,
Theorem~\ref{nelson:oppen:extension:thm} can be iterated to combine
theories $T_1, T_2, T_3, \ldots$ with the requisite properties.

Let us consider some examples of fragments of the reals that admit
quantifier-elimination, and are hence decidable. Note that to
eliminate quantifiers from any formula it suffices to be able to
eliminate a single existential quantifier, i.e.~transform a formula
$\ex x \ph$, where $\ph$ is quantifier-free, to an equivalent
quantifier-free formula. Since $\ex x (\ph \lor \psi)$ is equivalent
to $\ex x \ph \lor \ex x \psi$, we can always factor existential
quantifiers through a disjunction. In particular, since any
quantifier-free formula can be put in disjunctive normal form, it
suffices to eliminate existential quantifiers from conjunctions of
atomic formulas and their negations.  Also, since $\ex x (\ph \land
\psi)$ is equivalent to $\ex x \ph \land \psi$ when $x$ is not free in
$\psi$, we can factor out any formulas that do not involve $x$.
Furthermore, whenever we can prove $\fa x (\theta \lor \eta)$, $\ex x
\ph$ is equivalent to $\ex x (\ph \land \theta) \lor \ex x (\ph \land
\eta)$; so we can ``split across cases'' as necessary. We will use all
of these facts freely below.

\begin{proposition}
  The theory of $\la \RR, 0, 1, +, -, < \ra$ admits elimination of
  quantifiers, and hence is decidable.
\end{proposition}

This theory is commonly known as linear arithmetic, and is the same as
the theory of divisible ordered abelian groups. The universal fragment
coincides with that of the theory of ordered abelian groups. The
method of eliminating an existentially quantified variable implicit in
the proof is known as the \emph{Fourier-Motzkin procedure}.

\proof
  It is helpful to extend the language to include multiplication by
  rational coefficients, though we can view this as nothing more than
  a notational convenience: for example, if $n$ is a natural number,
  we can take $nx$ to abbreviate $x + x + \ldots + x$, and when $n, m,
  k, l$ are natural numbers with $m$ and $l$ nonzero we can take
  $(n / m) s = (k / l) t$ to abbreviate $nl s = km t$.
  
  Consider a sentence $\ex x \ph$, where $\ph$ is quantifier-free.
  Writing $s \neq t$ as $s < t \lor t < s$ and $s \not < t$ as $t < s
  \lor t = s$, we can assume without loss of generality that $\ph$ is
  a positive boolean combination of atomic formulas of the form $s =
  t$ and $s < t$. Putting $\ph$ in disjunctive normal form and
  factoring the existential quantifier though the disjunction we can
  assume $\ph$ is a conjunction of atomic formulas. Solving for $x$,
  we can express each of these in the form $x = s$, $x < s$, or $s <
  x$, where $s$ does not involve $x$ (atomic formulas that do not
  involve $x$ can be brought outside of the existential quantifier).

  If any of the conjuncts is of the form $x = s$, then $\ex x \ph(x)$
  is equivalent to $\ph(s)$, which is quantifier-free. So we are
  reduced to the case where $\ph$ is of the form $(\bigwedge_i s_i <
  x) \land (\bigwedge_j x < t_j)$. In that case, it is not hard to
  verify that $\ex x \ph$ is equivalent to $\bigwedge_{i,j} s_i <
  t_j$.\qed

For more on the Fourier-Motzkin procedure, see \cite{apt:03}. In
fact, more efficient elimination procedures are available, and are not
much more complicated; see
\cite{loos:weispfenning:93,weispfenning:88}.

\begin{proposition}
\label{mult:elim:prop}
The theory of $\la \RR, 0, 1, -1, \times, \div, < \ra$ with the
convention $x \div 0 = 0$ admits elimination of quantifiers, and hence
is decidable.
\end{proposition}

\proof
  Since $\la \RR^{\mathord{>}0}, 1, \times, \div, < \ra$ is isomorphic
  to $\la \RR, 0, +, -, < \ra$, the previous argument shows that the
  theory of this structure has quantifier-elimination. For the larger
  structure, consider $\ex x \ph$, where $\ph$ is quantifier-free. As
  above, we can assume $\ph$ is a conjunction of equalities and strict
  inequalities.  Introducing case splits we can assume that $\ph$
  determines which variables are positive, negative, or
  $0$. Temporarily replacing negative variables by their negations, we
  can further assume that $\ph$ implies that all the variables are
  positive. Bringing negation symbols to the front of each term, we
  are left with a conjunction of atomic formulas of the form $\pm s <
  \pm t$, where $s$ and $t$ are products of variables assumed to be
  positive.  But then $-s < t$ is equivalent to $\top$; $s < -t$ is
  equivalent to $\bot$; and $-s < -t$ is equivalent to $t <
  s$. Similarly, $-s = -t$ is equivalent to $s = t$, and both $s = -t$
  and $-s = t$ are equivalent to $\bot$. So, we are reduced to the
  case where all the variables are positive.\qed

\begin{proposition}
  The theory of $\la \RR, \exp, \ln, 0, 1, < \ra$, where
  $\exp(x) = e^x$ and $\ln(x) = 0$ for non-positive $x$, admits
  quantifier-elimination, and hence is decidable.
\end{proposition}

\proof
  Once again, we are reduced to the case of eliminating a quantifier
  of the form $\ex x \ph$ where $\ph$ is a conjunction of equalities
  and strict inequalities. Expressions of the form $\ln (\exp(s))$
  simplify to $s$, and across a case split of the form $s > 0 \lor s
  \leq 0$ an expression of the form $\exp(\ln(s))$ simplifies to $s$
  or $0$. Using the equivalences $s < t \liff \exp(s) < \exp(t)$ and
  introducing case splits as necessary, we are reduced to the case
  where $\ph$ is a conjunction of terms of the form $u < \exp^n(v)$,
  $u > \exp^n(v)$, and $u = \exp^n(v)$, where $u$ and $v$ are
  variables and $\exp^n(u)$ denotes $n$ applications of $\exp$ to $u$.
  If there is an equality using $x$, we can use that to eliminate the
  existential quantifier. Otherwise, for suitable $k$ we can arrange
  that $\ph$ is a conjunction of formulas of the form $s_i < \exp^k(x)$
  and $\exp^k(x) < t_j$, in which case $\ex x \ph$ is equivalent to
  $(\bigwedge_{i,j} s_i < t_j) \land (\bigwedge_j 0 < t_j)$.\qed

From Theorem~\ref{nelson:oppen:extension:thm} we have:

\begin{corollary}
\label{bad:decidability:thm}
The universal fragment of the union of the three theories above is
decidable. 
\end{corollary}

The decision procedure implicit in the proof of
Corollary~\ref{bad:decidability:thm} is, unfortunately, not very
useful. There is a sense in which is does too little, and another
sense in which it does too much.

A sense in which the procedure does too little is that the union of
the three theories is too weak. For example, it is not hard to show
(either using the interpolation theorem or a model-theoretic argument)
that the theory does not prove $\bar 2 \times \bar 2 = \bar 4$, where
$\bar 2$ abbreviates the term $1 + 1$, and $\bar 4$ abbreviates $1 + 1
+ 1 + 1$. Similarly, it fails to prove $x + x = \bar 2 x$. In the next
section, we will focus on the additive and multiplicative fragments of
the reals, and respond to this problem by augmenting the structures to
allow multiplication by arbitrary rational constants, or, more
generally, constants from a suitably computable subfield $F$ of the
reals. Unfortunately, this means that the two structures share a
language with infinitely many function symbols, and so the methods
described in the last section can no longer be used. We will have to
do a good deal of additional work to establish decidability in this
case.

A sense in which the algorithm implicit in the proof of
Corollary~\ref{bad:decidability:thm} does too much is that even in the
absence of the new multiplicative constants, it is inefficient: the
combination procedure relies on the fact that one can enumerate all
possible descriptions of equalities and inequalities between
variables, and, in general, the number of possibilities grows
exponentially. Our proof of decidability for the augmented theories
involves a reduction to the theory of real-closed fields, and so it
does not represent a practical advance either. In
Sections~\ref{avoiding:section}--\ref{extensions:section}, we will
address the issue of developing practical procedures that approximate
the theories we describe here.


\section{The theories $T[F]$}
\label{theories:section}

Let $F$ denote any subfield of the reals. Let $\Tadd[F]$ be the
theory of the real numbers for the language with symbols
\[
0, 1, +, -, <, \ldots, f_a, \ldots
\]
where for $a \in F$, $f_a$ denotes the function $f_a(x) = ax$. Let
$\Tmult[F]$ be the analogous theory for the language with symbols
\[
0, 1, \times, \div, <, \ldots, f_a, \ldots
\]
where $x \div y$ is interpreted as $0$ when $y = 0$. Our central
concern in this paper is the union of these two theories, $T[F] =
\Tadd[F] \cup \Tmult[F]$. It will also be useful to denote their
intersection, $\Tadd[F] \cap \Tmult[F]$, by $\Tcommon[F]$. It often
makes sense to restrict one's attention to computable subfields $F$ of
the real numbers; in particular, $\QQ$, the minimal such subfield,
is a natural choice. We will see below that, in a sense, the field of
real algebraic numbers $\AAA$ represents a maximal choice.
Intermediate choices are also possible; for example, one might
consider the smallest field containing $\QQ$ and closed under taking
roots of positive numbers. It should be clear that each $T[F]$ proves,
for example, $\bar 2 \times \bar 2 = \bar 4$ and $x + x = \bar 2x$.

We claim that the theories $T[F]$ are natural, and are sufficient to
justify many of the inferences that come up in ordinary mathematical
texts. The latter claim is an empirical one, however, and we will not
try to justify it here.

Each of $\Tcommon[F]$, $\Tadd[F]$, and $\Tmult[F]$ has
quantifier-elimination, and hence is complete. The elimination
procedures sketched in Section~\ref{fragments:section} can easily be
extended to $\Tadd[F]$ and $\Tmult[F]$, assuming the operations
on $F$ are computable, in which case these theories are decidable as
well. Similarly, a quantifier-elimination procedure for $\Tcommon[F]$
is easily obtained by extending the usual procedure for dense linear
orders without endpoints, so this theory is also complete, and
decidable when $F$ is computable.

Reflecting these elimination procedures yields complete
axiomatizations of the relevant theories. The theory $\Tcommon[F]$ is
axiomatized by the following:
\begin{enumerate}
\item $<$ is a dense linear order
\item $0 < 1$
\item $f_a(f_b(x)) = f_{ab}(x)$, for every $a, b \in F$
\item $f_0(x) = 0$, $f_1(x) = x$
\item $x < y \liff f_a(x) < f_a(y)$ for $0 < a \in F$
\item $x < y \liff f_a(x) > f_a(y)$ for $0 > a \in F$
\item $0 < x \limplies x < f_a(x)$ for $1 < a \in F$
\end{enumerate}
One obtains an axiomatization for $\Tadd[F]$ by adding the
following:
\begin{enumerate}
\item $0, +, <$ is an ordered abelian group
\item $x - y = z \liff x = y + z$
\item $f_a(x + y) = f_a(x) + f_a(y)$
\item $f_{a + b}(x) = f_a(x) + f_b(x)$
\end{enumerate}
Similarly, one obtains an axiomatization of $\Tmult[F]$ by adding the
following to $\Tcommon[F]$:
\begin{enumerate}
\item $1, \times, <$ is a divisible ordered abelian group on the
  positive elements
\item $x / y = z \liff (y = 0 \land z = 0) \lor x = y z$
\item $f_a(xy) = f_a(x)y$
\end{enumerate}

In Sections \ref{models:section}--\ref{undecidability:two:section}, we
will prove undecidability results for fragments of $T[F]$. We
will find it useful to work with the following alternative system,
$T[F]^*$, based on the symbols $0,1,+,\times,<$ together with
constant symbols $c_a$ for $a \in F$. The axioms of $T[F]^*$ fall
naturally into four groups:
\begin{enumerate}
\item $0,+,<$ is an ordered abelian group
\item $1,\times,<$ is a divisible ordered abelian group on the
  positive elements
\item 
\begin{enumerate}
\item $c_{a+b} = c_a + c_b$, for $a,b \in F$
\item $c_{ab} = c_a \times c_b$, for $a,b \in F$
\item $0 < c_a$ for $0 < a$, $a \in F$
\end{enumerate}
\item
\begin{enumerate}
\item $c_{a+b} \times x = (c_a \times x) + (c_b \times x)$, for $a,b \in F$
\item $c_a \times (x+y) = (c_a \times x) + (c_a \times y)$, for $a \in F$
\end{enumerate} 
\end{enumerate}

Note that the extra symbols in the language of $T[F]$ are easily
definable in $T[F]^*$. It is straightforward to verify the following.

\begin{lemma}
\label{und:one:one} 
Let $\ph$ be a formula in the language of $T[F]$ without
$-,\div$. Let $\ph^*$ be the result of replacing
each occurrence of $f_a(t)$ with $c_a \times t$, inductively, from
innermost to outermost. Then $\ph$ is provable in $T[F]$ if and only
if $\ph^*$ is provable in $T[F]^*$.\qed
\end{lemma}

\begin{lemma}
\label{und:one:two}
Let $\ph$ be a formula in the language of $T[F]^*$. Let $\ph'$ be the
result of replacing each occurrence of $c_a$ with $f_a(1)$. Then $\ph$
is provable in $T[F]^*$ if and only if $\ph'$ is provable in $T[F]$.\qed
\end{lemma}

\begin{theorem}
\label{und:one:three}
$T[F]$ and $T[F]^*$ prove the same sentences involving only the symbols
$0,1,+,\times,<$.\qed
\end{theorem}

Below we will call the symbols $f_a$ the \emph{auxiliary function
  symbols} and the symbols $c_a$ the \emph{auxiliary constant
  symbols}. For readability, we will write $a x$ instead of
$f_a(x)$ or $c_a x$ when the context makes the meaning clear.

The following shows that as far as provability of formulas in the
language of real closed fields is concerned, there is never a need to
go beyond the real algebraic numbers in choosing $F$.

\begin{theorem}
\label{alg:maximal:thm}
$T[\RR]$ is a conservative extension of $T[\AAA]$.
\end{theorem}

\proof
  Since $\div$ and $-$ are definable in terms of the other symbols of
  $T[F]$, we can focus on sentences in which these symbols do not
  occur, and use Theorem~\ref{und:one:three}.
  
  Let $d$ be a proof of a sentence $\ph$ in $T[\RR]^*$, where $\ph$ is
  in the language of $T[\AAA]^*$. Assign variables $\vec y$ to the
  auxiliary constant symbols occurring in $\ph$, and let $\psi(\vec y)$
  define the corresponding real algebraic numbers in the language of
  real closed fields. Assign variables $\vec z$ to all the additional
  auxiliary constant symbols occurring in $d$, and let $\theta(\vec
  y,\vec z)$ be the conjunction of all the axioms of $T[\RR]$ used in
  $d$, with the constants replaced by the corresponding variables. The
  assertion $\ex {\vec y, \vec z} (\theta(\vec y, \vec z) \land
  \psi(\vec y))$ is true of the real numbers, and so, by transfer
  (i.e.~the completeness of the theory of real closed fields, of which
  both the reals the real algebraic numbers are a model), it is true
  of $\AAA$ as well. Let $\vec a, \vec b$ be real algebraic numbers
  witnessing the existential quantifiers. Because $\psi(\vec a)$
  determines $\vec a$ uniquely, $\vec a$ corresponds to the original
  auxiliary constant symbols in $\ph$. Thus we have the even stronger
  result that $d$ can be interpreted as a proof in $T[\AAA]^*$, taking
  the constant symbols to denote $\vec a, \vec b$.\qed

This argument shows, more generally, that to prove a sentence with
auxiliary function symbols $f_{a_1}, \ldots, f_{a_n}$, there is no
need to go beyond the real algebraic closure of $\{ a_1, \ldots, a_n
\}$.


\section{Examples}
\label{examples:section}

To provide a better feel for the theories $T[F]$, in this section we
consider some theorems that clarify their strength. The first theorem
provides a lower bound by showing that a decision procedure for the
universal fragment of any $T[F]$ implies a decision procedure for the
existence of roots of a multivariate polynomial on the unit cube.

\begin{theorem}
  Let $F$ be any subfield of the real numbers, and let
  $f(x_1,\ldots,x_k)$ be a multivariate polynomial with coefficients
  in $F$. Let $I = [0,1]^k$ be the compact $k$-dimensional unit cube.
  Then $f$ is nonzero on $I$ if and only if $T[F]$ proves that fact.
\end{theorem}

\proof
  The ``if'' direction follows from the fact that the axioms of $T[F]$
  are true of the real numbers. On the other hand, by the intermediate
  value theorem, if a polynomial function $f$ is nonzero on $I$, then
  it is either strictly positive or strictly negative on $I$. So it
  suffices to show that if $f$ is strictly positive on $I$, then
  $T[F]$ proves that this is the case.
  
  Suppose $f(\vec x) = \sum_{i < n} t_i(\vec x)$, where each $t_i$ is
  a monomial in $x_1, \ldots, x_l$ with a coefficient in $F$, and
  suppose $f$ is strictly positive on $I$. Given a point $\la a_1,
  \ldots, a_k \ra$ in $I$, let $r_{\vec a} = f(\vec a) > 0$, and for
  each $i$, let $r_{\vec a,i} = t_i(\vec a)$. By continuity, we can
  find an open neighborhood $U_{\vec a}$ of $\vec a$, such that for
  each $\vec b \in U_{\vec a}$, $t_i(\vec b) > r_{i,\vec a} - r_{\vec
    a} / 3n$. Shrinking $U_{\vec a}$ if necessary, we can assume that
  $U_{\vec a}$ is a product of open intervals with rational endpoints.

By compactness, $U$ is covered by a finite set of these open
neighborhoods, say $U_{\vec a_1}, \ldots, U_{\vec a_m}$. Then:
\begin{enumerate}
\item $T[F]$ proves $\fa {\vec x} (\vec x \in I \limplies \vec x \in
  U_{\vec a_1} \lor \ldots \lor U_{\vec a_m})$. In fact, this can
  be proved by $\Tcommon[F]$, since it is purely a property of the
  ordering on the rational numbers.
\item For each $j<m$ and $i<n$, $\Tmult[F]$ proves $\vec x \in
  U_{\vec a_j} \limplies t_i(\vec x) > q_{i,j}$, where $q_{i,j}$ is
  any rational number less than $r_{\vec a_j,i} - r_{\vec a_j} /
  3n$ and greater than $r_{\vec a_j,i} - r_{{\vec a}_j} / 2n$.
\item Using these lower bounds, for each $j < m$, $\Tadd[F]$ can prove
  $\vec x \in U_{\vec a_j} \limplies f(\vec x) > \sum_{i < n}
  q_{i,j}$.
\end{enumerate}
The result follows from the fact that in the last claim,
\[
\sum_{i < n} q_{i,j} > \sum_{i < n} (r_{\vec a_j,i} - r_{\vec a_j}
/ 2n) = r_{\vec a_j} - r_{\vec a_j} / 2 = r_{\vec a_j} / 2 > 0.
\]
This completes the proof.\qed

As an example of something $T[F]$ \emph{cannot} do, consider the
inequality $x^2 - 2x + 1 \geq 0$. That this is generally valid is
clear from writing $x^2 - 2x + 1 = (x - 1)^2$, but this equality is a
consequence of distributivity, which is not available in $T[F]$. In
fact, we have:

\begin{theorem}
\label{upper:bound:one:thm}
For any $F$, $T[F]$ proves $\fa x (x^2 - 2x + 1 \geq \varepsilon)$ if
and only if $\varepsilon < 0$. In particular, $T[F]$ does not prove
$\fa x (x^2 - 2x + 1 \geq 0)$.
\end{theorem}

Moreover, proofs of $\fa x (x^2 - 2x + 1 \geq \varepsilon)$ in $T[F]$
necessarily get longer as $\varepsilon$ approaches $0$, and the
results that follow provide explicit lower bounds. Focusing on the
domain of the function $x^2 - 2x + 1$ instead of the range, we also
have:

\begin{theorem}
\label{upper:bound:two:thm}
For any $F$,
\begin{enumerate}
\item $T[F]$ proves $\fa x (x \leq r \limplies x^2 - 2x + 1 \geq 0)$
  if and only if $r < 1$.
\item $T[F]$ proves $\fa x (x \geq r \limplies x^2 - 2x + 1 \geq 0)$
  if and only if $r > 1$.
\end{enumerate}
\end{theorem}

Theorem~\ref{upper:bound:two:thm} implies
Theorem~\ref{upper:bound:one:thm}. Assuming $x \in [1 - \delta, 1 +
\delta]$ for a small rational constant $\delta$, $T[F]$ can easily
show $x^2 \geq 1 - 2\delta + \delta^2$ and $2x \leq 2 + 2\delta$, and
hence $x^2 - 2x + 1 \geq - 4 \delta + \delta^2 \geq - 4 \delta$. So,
taking $r$ to be $1 - \delta$ and $1 + \delta$, respectively, in the
two clauses Theorem~\ref{upper:bound:two:thm}, we have the ``if''
direction of Theorem~\ref{upper:bound:one:thm}. But the ``only if''
direction is a consequence of the fact that $T[F]$ does not prove
$\fa x (x^2 - 2x + 1 \geq 0)$, which is immediate from
Theorem~\ref{upper:bound:two:thm}.

The two clauses of Theorem~\ref{upper:bound:two:thm} are proved in a
similar way, and so we will only prove the first. Since $T[F]$ easily
proves $x < 0 \limplies x^2 - 2x + 1 \geq 0$, we can replace the first
statement in Theorem~\ref{upper:bound:two:thm} by $\fa x (0 \leq x
\leq r \limplies x^2 \geq 2x - 1)$. $T[F]$ proves this if and only if
it refutes the set of formulas
\[
\{ 0 \leq x, x \leq r, u = x^2, u < 2x - 1 \}.
\]
Recall that this happens if and only if there is an interpolant,
$\theta$, in disjunctive normal form, such that
\begin{equation}
\label{upper:bound:eq:one}
\Tmult[F] \cup \{ 0 \leq x, x \leq r, u = x^2 \} \proves \theta
\end{equation}
and
\begin{equation}
\label{upper:bound:eq:two}
\Tadd[F] \cup \{ u < 2x - 1 \} \cup \theta \proves \bot.
\end{equation}
So it suffices to show:

\begin{theorem}
\label{upper:bound:three:thm}
There is a $DNF$ formula $\theta$ with at most $n$ disjuncts
satisfying (\ref{upper:bound:eq:one}) and (\ref{upper:bound:eq:two})
if and only if $r <= n / (n + 1)$.
\end{theorem}

\proof
  We will first show that if $\theta$ has $n$ disjuncts and satisfies
  (\ref{upper:bound:eq:one}) and (\ref{upper:bound:eq:two}) then $r
  \leq n / (n + 1)$. We will then show that, in fact, for $r = n / (n
  + 1)$ such a $\theta$ exists.

Write $\theta = \theta_1 \lor \ldots \lor 
\theta_n$, where each $\theta_i$ is a conjunction of literals
involving only $x$ and $u$. It is
not hard to see that each $\theta_i$ is equivalent to a conjunction of
literals of the form 
\[
a \triangleleft x \triangleleft b \land c \triangleleft u \triangleleft
d \land e x \triangleleft u \triangleleft f x
\]
where each $\triangleleft$ is either $<$ or $\leq$ (and some of the
conjuncts may be absent). $\Tmult[F]
\cup \{ 0 \leq x, x \leq r, u = x^2 \}$ proves this equivalent to a
conjunction of the form
\begin{equation}
\label{eq:a}
a \triangleleft x \triangleleft b \land a^2 \triangleleft u
\triangleleft b^2 \land a x \triangleleft u \triangleleft b x
\end{equation}
for some $a, b$ in $[0,1]$, and from the point of view of $\Tadd[F]
\cup \{ 2x - 1 < u \}$, each of these disjuncts is no weaker than the
original. Thus it suffices to prove the claim for interpolants that
are of the form (\ref{eq:a}).

Now, $\Tadd[F] \cup \{ u < 2x - 1 \}$ refutes $\theta$ if and only if it
refutes each disjunct. Thus the following lemma is crucial to our
analysis. 


\begin{lemma}
\label{upper:bound:lemma}
For $a, b$ in $[0,1)$, $\Tadd[F] \cup \{ 2x - 1 < u \}$ refutes
(\ref{eq:a}), for any versions of the relation $\triangleleft$, if and
only if $b \leq 1 / (2 - a)$.
\end{lemma}

\proof
  If $b < a$, $\Tadd[F] \cup \{ 2x - 1 < u \}$ easily refutes
  (\ref{eq:a}), and $b \leq 1 / (2 - a)$ holds. So it suffices to
  consider the case $a \leq b$.

We need only work through the Fourier-Motzkin procedure by hand.
Eliminating $u$, we obtain the inequalities $a^2 < 2x - 1$ and $ax <
2x - 1$. (Note that we get strict inequality, whether the initial
$\triangleleft$'s are strict inequalities or not.) Solving for $x$, we
obtain $(a^2 + 1) / 2 < x$ and $1 / (2 - a) < x$. Eliminating $x$, we
get $(a^2 + 1) / 2 < b$ and $1 / (2 - a) < b$. This yields a
contradiction if and only if $b$ is less than or equal to the minimum
of $(a^2 + 1) / 2$ and $1 / (2 - a)$. A calculation shows
that the latter is always smaller for $a \in [0,1)$, so we have the
desired conclusion.\qed

We can now finish off the proof of
Theorem~\ref{upper:bound:three:thm}. Suppose $\Tmult[F] \cup \{ 0
\leq x, x \leq r, u = x^2 \}$ proves a disjunction $\theta_1 \lor
\ldots \lor \theta_n$ with each $\theta_i$ of the form (\ref{eq:a})
for some $a_i$ and $b_i$. If any of the intervals $(a_i,b_i)$ overlap,
we can strengthen some disjuncts (and eliminate redundant ones) and
obtain an equivalent interpolant where the intervals $(a_i,b_i)$ are
disjoint and are listed so that for each $i$, $a_i < a_{i+1}$. On the
other hand, $\Tadd[F] \cup \{ 2x - 1 < u \}$ refutes $\theta$ if and
only if it refutes each $\theta_i$, and if this is the case, it is
certainly true for any $\theta_i'$ such that $\Tadd[F] \cup \{ 2x - 1
< u \}$ proves $\theta'_i \limplies \theta_i$. Thus, from the point of
view of proving the ``only if'' direction of the theorem, we may
assume, without loss of generality, that $\theta$ is a disjunction of
formulas of the form (\ref{eq:a}), and the intervals $(a_i,b_i)$
corresponding to the $a$ and $b$ in each $\theta_i$ are increasing and
disjoint. 

But then it is clear that $\Tmult[F] \cup \{ 0 \leq x, x \leq r, u = x^2
\}$ proves $\theta_1 \lor \ldots \lor \theta_n$ if and only if
\begin{enumerate}
\item $a_0 = 0$, 
\item $b_i = a_{i+1}$, for each $i<n$,
\item $a_n = r$,
\end{enumerate}
and the $\triangleleft$'s are chosen suitably.
Lemma~\ref{upper:bound:lemma} guarantees that for each $i$, $a_{i+1}
\leq 1 / (2 - a_i)$. The largest possible value of $r$ occurs when the
inequality is replaced by an equality $a_{i+1} = 1 / (2 - a_i)$, and a
calculation shows that in that case, $a_i = i / (i + 1)$ for each $i
\leq n$.

This proves the ``only if'' direction of the theorem, establishing an
upper bound on the possible values of $r$. But the proof in fact
yields an interpolant that shows that the upper bound can be obtained:
if each $\theta_i$ is the formula
\[
a_i \leq x \leq a_{i+1} \land a_i^2 \leq u \leq a_{i+1}^2
\land a_i x \leq u \leq a_{i+1} x
\]
with $a_i = i / (i + 1)$, then $\Tmult[F] \cup \{ 0 \leq x, x \leq r,
u = x^2 \}$ proves $\theta_1 \lor \ldots \lor \theta_n$, and $\Tadd[F]
\cup \{ 2 x - 1 < u \}$ refutes each $\theta_i$.\qed


\section{Provability of a universal sentence in $T[F]$}
\label{provability:section}

In this section, we will provide various characterizations of
provability of a universal sentence in $T[F]$. These will be used in
Section~\ref{decidability:section} to establish our decidability
results.

By Proposition~\ref{separating:variables:prop}, if $\ph$ is a
universal sentence in the language of some $T[F]$, $\ph$ is equivalent
to a formula of the form $\fa {\vec x} (\phadd(\vec x) \lor
\phmult(\vec x))$, where $\phadd$ and $\phmult$ are in the language of
$\Tadd[F]$ and $\Tmult[F]$, respectively.

\begin{proposition}
  Let $\ph \equiv \fa{\vec x} (\phadd(\vec x) \lor \phmult(\vec x))$
  be as above. Then the following are equivalent:
\begin{enumerate}
\item $T[F]$ proves $\ph$.
\item There is a quantifier-free formula $\theta(\vec x)$ in the
  language $\Tcommon[F]$ such that $\Tadd[F] \cup
  \{ \theta(\vec x) \} \proves \phadd(\vec x)$ and $\Tmult[F] \cup \{
  \lnot \theta (\vec x) \} \proves \phmult(\vec x)$.
\item There is a quantifier-free formula $\theta(\vec x)$ in the
  language $\Tcommon[F]$ such that
\[
\fa {\vec x} (\theta(\vec x) \limplies \phadd(\vec x)) \quad
\mbox{and} \quad \fa {\vec x} (\lnot \theta(\vec x) \limplies
\phmult(\vec x))
\]
hold of the real numbers, with the intended interpretation of the
auxiliary function symbols.
\end{enumerate}
\end{proposition}

\proof
  If 2 holds, then clearly $T[F]$ proves $\phadd(\vec x) \lor
  \phmult(\vec x)$. Thus 2 implies 1. Conversely, if $T[F]$ proves
  $\ph$, it proves $\lnot \phmult(\vec x) \limplies \phadd(\vec x)$.
  Treating $\vec x$ as new constants and applying the Craig
  interpolation lemma, we get an interpolant $\theta(\vec x)$ in the
  language of $\Tcommon[F]$ satisfying the conclusion of 2. Since
  $\Tcommon[F]$ has quantifier-elimination, we can assume without loss
  of generality that $\theta(\vec x)$ is quantifier-free.

The equivalence of 2 and 3 follows easily from the fact that each of
$\Tadd[F]$ and $\Tmult[F]$ is a complete theory that holds of the
reals numbers with the intended interpretation of the auxiliary function
symbols.\qed

From a model-theoretic perspective, it is useful to replace
provability by nonexistence of a countermodel. When we say
$\Gamma(\vec x)$ is a \emph{type} over a theory $T$, we mean that
$\Gamma$ is a set of formulas in the language of $T$, involving only
the free variables $\vec x$, such that $\Gamma$ is consistent with $T$.
Saying $\Gamma(\vec x)$ is a \emph{complete type} means that for every
formula $\psi(\vec x)$, either $\psi(\vec x)$ or $\lnot \psi(\vec x)$
is in $\Gamma(\vec x)$.

\begin{proposition}
\label{model:theoretic:equivalences:prop}
Let $\ph \equiv \fa{\vec x} (\phadd(\vec x) \lor \phmult(\vec x))$ be
as above. Then the following are equivalent:
\begin{enumerate}
\item $T[F]$ does not prove $\ph$.
\item $T[F] \cup \{ \lnot \ph \}$ is consistent.
\item The union of $\Tadd[F] \cup \{ \lnot \phadd(\vec x) \}$
  and $\Tmult[F] \cup \{ \lnot \phmult(\vec x) \}$ is consistent.
\item There is a complete type $\Gamma(\vec x)$ over 
  $\Tcommon[F]$ such that
\[
\Tadd[F] \cup \Gamma(\vec x) \cup \{ \lnot \phadd(\vec x) \} \quad
\mbox{and} \quad \Tmult[F] \cup \Gamma(\vec x) \cup \{ \lnot
\phmult(\vec x) \}
\]
are both consistent.
\item There is a complete type $\Gamma(\vec x)$ over $\Tcommon[F]$
  such that for every finite $\Gamma'(\vec x) \subseteq \Gamma(\vec
  x)$,
\[
\Tadd[F] \proves \ex {\vec x} (\bigwedge \Gamma'(\vec x) \land \lnot
\phadd(\vec x))\]
and
\[
\Tmult[F] \proves \ex {\vec x}
(\bigwedge \Gamma'(\vec x) \land \lnot \phmult(\vec x)).
\]
\item There is a complete type $\Gamma(\vec x)$ over $\Tcommon[F]$
  such that for every finite $\Gamma'(\vec x) \subseteq \Gamma(\vec
  x)$, there are real numbers $\vec x$ and $\vec y$ satisfying
\[
\Gamma'(\vec x) \land \lnot \phadd(\vec x) \land \Gamma'(\vec y) \land
\lnot \phmult(\vec y).
\]
\end{enumerate}
\end{proposition}

\proof
  In light of the soundness and completeness of first-order logic, 1
  is just a restatement of 2, and the equivalence with 3 follows from
  the definition of $\ph$ in terms of $\phadd$ and $\phmult$. The
  equivalence of 3 with 4 follows by the Robinson joint consistency
  theorem, or, equivalently, from the Craig interpolation theorem,
  using compactness.
  
  That statement 4 implies statement 5 follows from the fact that
  $\Tadd[F]$ and $\Tmult[F]$ are both complete theories; for example,
  $\Tadd[F] \cup \Gamma'(\vec x) \cup \{ \lnot \phadd(\vec x) \} $ is
  consistent if and only if $\Tadd[F]$ proves $\ex {\vec x} (\bigwedge
  \Gamma'(\vec x) \land \lnot \phadd(\vec x))$. The converse is immediate.
  
  The equivalence of 5 and 6 follows from the fact that each of
  $\Tadd[F]$ and $\Tmult[F]$ is the theory of the real numbers in the
  respective languages.\qed

Note that the equivalence of 1--4 holds, in general, for any two
theories. The equivalence with 5 relies only on the fact that
$\Tadd[F]$ and $\Tmult[F]$ are complete, and the equivalence with 6
relies only on the additional fact that they are satisfied by the
reals.

Statement 6 provides a nice characterization of provability in $T[F]$.
A universal sentence $\ph$ is true of the reals if and only if every
sequence $\vec x$ of reals satisfies either $\phadd(\vec x)$ or
$\phmult(\vec x)$. But a universal sentence $\ph$ is provable in
$T[F]$ if and only if for every complete type $\Gamma(\vec x)$ in the
language of $\Tcommon[F]$, there is a finite subset $\Gamma'(\vec x)$
such that either
\[
\fa {\vec x} (\bigwedge \Gamma'(\vec x) \limplies \phadd(\vec x))
\quad \mbox{or} \quad \fa {\vec x} (\bigwedge
\Gamma'(\vec x) \limplies \phmult(\vec x))
\]
holds in the reals. In particular, this has to hold whenever
$\Gamma(\vec x)$ is the type corresponding to a sequence of real
numbers; but we will see below that there are types in the language of
$\Tcommon[F]$ that are not of this form. Thus, provability in $T[F]$
imposes a stronger requirement.

In the remainder of this section, we consider various representations
of the quantifier-free formulas $\phadd(\vec x)$, $\phmult(\vec x)$,
and the possible interpolants $\theta(\vec x)$. We also consider
representations of the types $\Gamma(\vec x)$. The former will be
relevant to the discussion of heuristic algorithms in
Sections~\ref{avoiding:section}--\ref{extensions:section}, whereas the
latter will be used in our decidability proofs in
Section~\ref{decidability:section}. 

Let $\ph \equiv \fa{\vec x} (\phadd(\vec x) \lor \phmult(\vec x))$ be
as above.  Since $\fa y \psi(y)$ is equivalent to $\fa {y > 0} \psi(y)
\land \psi(0) \land \fa {y > 0} \psi(-y)$, as in the proof of
Proposition~\ref{mult:elim:prop}, any universal sentence $\ph$ is
equivalent to a conjunction of formulas of the form $\fa {\vec x > 0}
(\phadd(\vec x) \lor \phmult(\vec x))$. We can absorb the condition
$\vec x > 0$ into both $\phadd(\vec x)$ and $\phmult(\vec x)$. By
adding a new variable if necessary, we can also assume that each
includes the condition $x_1 = 1$, and it will be notationally
convenient to do so. Thus, for the rest of this section, we will
assume that $\ph$ is a universal formula of the form $\fa {\vec x}
(\phadd(\vec x) \lor \phmult(\vec x))$ where $\phadd(\vec x)$ and
$\phmult(\vec x)$ are quantifier-free in the language of $\Tadd[F]$
and $\Tmult[F]$, respectively, and $\lnot \phadd(\vec x)$ and $\lnot
\phmult(\vec x)$ each implies $\vec x > 0$ and $x_1 = 1$.  The
question as to the decidability of the universal fragment of $T[F]$
reduces to the question as to whether one can determine whether $T[F]$
proves a sentence of this form. Let $\Delta(\vec x)$ be the set $\{
\vec x > 0, x_1 = 1 \}$.

\begin{proposition}
\label{interpolant:prop}
  Under hypotheses $\Delta(\vec x)$, a quantifier-free formula in the
  language of $\Tcommon[F]$ can be put in any of the following forms:
\begin{enumerate}
\item a conjunction of disjunctions of atomic formulas of the form $x_i < a
  x_j$ or $x_i \leq a x_j$, with $a > 0$. 
\item a conjunction of disjunctions of atomic formulas of the form $x_i
  < a x_j$, with $a > 0$, or of the form $x_i = a x_j$ with $a > 0$
  and $i < j$.
\item either 1 or 2, with ``conjunction'' and ``disjunction''
  switched.
\end{enumerate}
\end{proposition}

\proof
  Let $\theta$ be quantifier-free. First, put $\theta$ in
  negation-normal form, so that it is built up from atomic formulas
  and negations of atomic formulas using $\land$ and $\lor$. Replace
  $s \not < t$ by $t \leq s$, replace $s \not\leq t$ by $t < s$, and
  replace $s \neq t$ by $s < t \lor t < s$. As a result,all atomic
  literals occur positively. One can further eliminate either $s \leq
  t$ in favor of $s < t \lor s = t$, or one can eliminate $s = t$ in
  favor of $s \leq t \land t \leq s$. The resulting formula can then
  be put in either disjunctive or conjunctive normal form, without
  introducing negations.
  
  In the end, all the atomic formulas are of the form $a x_i < b x_j$,
  $a x_i \leq b x_j$, or $a x_i = b x_j$. Dividing through by $b$ (and
  reversing an inequality when $b$ is negative), we can assume that in
  each case $b = 1$. With the assumptions in $\Delta$, each atomic
  formula in which $a$ is negative can be replaced by either $\top$ or
  $\bot$. Then inequalities $a x_i < x_j$ (resp.~$a x_i \leq x_j$) can
  be expressed as $x_i < (1 / a) x_j$ (resp.~$x_i \leq (1 / a) x_j$),
  as necessary, and equalities $x_j = a x_i$ can be rewritten $x_i =
  (1 / a) x_j$ when $i < j$.\qed

Such normal forms can be useful in reducing the problem of proof
search to restricted cases. From an implementation point of view, not
all these reductions are wise, however; for example, using case splits
to ensure that the $x$'s are all positive or to eliminate $s \leq t$
in favor of $s < t$ or $s = t$ can result in an exponential blowup. In
the absence of sign information, the normal forms are more
complicated. For example, although $x_2 > 2 x_3$ can be expressed as
$x_3 < (1/2)x_2$, $x_2 > -x_3$ cannot be expressed in the form $x_i <
ax_j$. Also, in the absence of sign information, neither of $x_2 <
x_3$ and $x_2 < 2x_3$ implies the other. In that case, one has to
consider normal forms with atomic formulas from among $x_i < ax_j$,
$x_i \leq a x_j$, $x_i > a x_j$, and $x_i \geq a x_j$. A little thought
shows that in a single conjunction or disjunction, for each pair $i,
j$, no more than two such formulas are needed; see also the proof of
Proposition~\ref{best:add:prop}.

We can similarly classify the complete types over $\Tcommon[F]$. Let
$\Gamma(\vec x) \supseteq \Delta(\vec x)$ be such a type. Since
$\Tcommon[F]$ has quantifier elimination, $\Gamma$ is determined by
the atomic formulas that it contains. Hence it is also determined by
its subsets $\Gamma_{i, j}(x_i,x_j)$, with $i < j$, where
$\Gamma_{i,j}$ consists of the atomic formulas involving both $x_i$
and $x_j$. If $\Gamma_{i, j}$ contains a formula of the form $x_i = a
x_j$, that determines the set $\Gamma_{i, j}$ uniquely. We denote this
type by $\Gamma_{x_i/x_j = a}$.  Otherwise, $\Gamma_{i,j}$ contains
the formula $x_i \neq a x_j$ for every $a$ in $F$, and so
$\Gamma_{i,j}$ is determined by the set of elements $a$ such that
$\Gamma_{i, j}$ contains the formula $x_i < a x_j$. This set is a
downwards-closed subset of the positive part of $F$; think of it as
the set of $a$ such that $x_i / x_j < a$.  If this set is empty, that
determines $\Gamma_{i,j}$ uniquely, and we denote the corresponding
type $\Gamma_{x_i / x_j \approx \infty}$.  Otherwise, the set has a
greatest lower bound in the real numbers, say, $r$. If $r$ is not an
element of $F$, then $\Gamma_{i,j}$ contains $x_i < a x_j$ exactly
when $r < a$, and this determines $\Gamma_{i,j}$ exactly; we denote
the resulting type by $\Gamma_{x_i / x_j \approx r}$. If, on the other
hand, $r$ is an element $a$ of $F$, there are two possibilities:
$\Gamma_{i,j}$ contains the formula $x_i < a x_j$, or it does not (in
which case it contains the formula $x_j < (1 / a) x_i$). Denote the
first type by $\Gamma_{x_i/x_j \approx a^-}$, and denote the second by
$\Gamma_{x_i/x_j \approx a^+}$.

In short, we have shown the following:

\begin{proposition}
\label{types:prop}
Let $\Gamma(\vec x)$ be any complete type over $\Tcommon[F]$
that includes $\Delta(\vec x)$. Then for each $i < j$, $\Gamma$
includes exactly one of the following:
\begin{enumerate}
\item $\Gamma_{x_i / x_j = a}$, for some $a$ in $F$
\item $\Gamma_{x_i / x_j \approx r}$, for some $r$ in $\RR \setminus F$
\item $\Gamma_{x_i /x_j \approx \infty}$
\item $\Gamma_{x_i / x_j \approx a^-}$, for some $a$ in $F$
\item $\Gamma_{x_i / x_j \approx a^+}$, for some $a$ in $F$
\end{enumerate}
These data determine $\Gamma$ uniquely.
\end{proposition}

Note that not every collection of sets $\Gamma_{x_i/x_j}$ determines a
consistent type over $\Tcommon[F]$; for example, the sets
$\Gamma_{x_1/x_2 = 2}$, $\Gamma_{x_2/x_3 = 2}$, and $\Gamma_{x_1/x_3 =
  2}$ are jointly inconsistent.

In the next section, we will combine the analysis given by
Proposition~\ref{types:prop}, together with equivalence 6 of
Proposition~\ref{model:theoretic:equivalences:prop}, to show that,
with general conditions on $F$, the universal fragment of $T[F]$ is
decidable.


\section{Decidability}
\label{decidability:section}

Let $\ph \equiv \fa {\vec x} (\phadd(\vec x) \lor \phmult(\vec x))$ be
as in the previous section, so that $\phadd$ and $\phmult$ are
quantifier-free formulas in the language of $\Tadd[F]$ and $\Tmult[F]$
respectively, and each of $\lnot \phadd(\vec x)$ and $\lnot
\phmult(\vec x)$ implies $\vec x > 0$ and $x_1 = 1$. We have seen that
the decidability of the universal fragment of $T[F]$ reduces to the
problem of determining whether $T[F]$ proves a formula $\ph$ of this
sort; and that $T[F]$ does \emph{not} prove such a $\ph$ if and only
if
\begin{quote}
there is a complete type
$\Gamma(\vec x)$ over $\Tcommon[F]$ such that for every
finite $\Gamma'(\vec x) \subseteq \Gamma(\vec x)$, the sentence
\[
\ex {\vec x} (\bigwedge \Gamma'(\vec x) \land \lnot \phadd(\vec x)) \land
\ex {\vec x} (\bigwedge \Gamma'(\vec x) \land \lnot \phmult(\vec x))
\]
is true of the real numbers. 
\end{quote}
Call this the ``consistency criterion for $\lnot \ph$.'' We also have
a complete classification of the relevant types $\Gamma(\vec x)$. In
this section, we will use the latter to show that when $F$ is a
computable subfield of $\RR$ and membership of a real algebraic number
in $F$ is decidable, the consistency criterion for $\lnot \ph$ is
decidable.

Fix $\ph$ and $F$, and hence $\phadd(\vec x)$ and $\phmult(\vec x)$.
If $\Gamma(\vec x)$ is any set of atomic formulas in the language of
$\Tcommon[F]$ involving the variables $\vec x$ and $i < j$, let
$\Gamma_{i,j}$ denote the set of formulas in $\Gamma$ involving $x_i$
and $x_j$. Let $\mathcal S$ be the collection of sets $\Gamma$ such
that for each $i < j$, $\Gamma_{i,j}$ is one of the types described in
Proposition~\ref{types:prop}. Since each such $\Gamma$ consistent with
$\Tcommon[F]$ uniquely determines the complete type that extends it,
we can replace ``complete type $\Gamma(\vec x)$ over $\Tcommon[F]$''
by ``$\Gamma \in \mathcal S$'' in the consistency criterion for $\lnot
\ph$.

We now show that we can modify the collection of sets $\mathcal S$ to
avoid the restrictions ``$a \in F$'' in the clauses of
Proposition~\ref{types:prop}. To do so, we consider types in the
larger language, $\Tcommon[\RR]$. Let the types $\hat \Gamma_{x_i /
  x_j = a}$, $\hat \Gamma_{x_i / x_j \approx r}$, $\hat \Gamma_{x_i
  /x_j \approx \infty}$, $\hat \Gamma_{x_i / x_j \approx a^-}$, and
$\hat \Gamma_{x_i / x_j \approx a^+}$ be defined as in the paragraph
before Proposition~\ref{types:prop}, except with respect to the
language of $\Tcommon[\RR]$. Let $\hat {\mathcal S}$ be the sets $\hat
\Gamma$ of atomic formulas in $\Tcommon[\RR]$ such that for each $i <
j$, $\hat \Gamma_{i,j}$ is one of the following:
\begin{enumerate}
\item $\hat \Gamma_{x_i / x_j = a}$, for some $a$ in $\RR$
\item $\hat \Gamma_{x_i / x_j \approx r}$, for some $r$ in $\RR \setminus F$
\item $\hat \Gamma_{x_i /x_j \approx \infty}$
\item $\hat \Gamma_{x_i / x_j \approx a^-}$, for some $a$ in $\RR$
\item $\hat \Gamma_{x_i / x_j \approx a^+}$, for some $a$ in $\RR$
\end{enumerate}
Note that we have replaced ``$a \in F$'' by ``$a \in \RR$'' in the
first item and in the last two items, but we have left $\RR \setminus
F$ alone in the second item.
\begin{lemma}
  The consistency criterion for $\lnot \ph$ is satisfied by a set
  $\Gamma \in {\mathcal S}$ if and only if it is satisfied by a set
  $\hat \Gamma \in \hat {\mathcal S}$.
\end{lemma}

\proof
  Suppose the consistency criterion is satisfied by some $\Gamma \in
  \mathcal S$. It is easy to check that it is then satisfied by the
  corresponding set $\hat \Gamma \in \hat {\mathcal S}$.

  In the other direction, note that if $a$ is in $\RR \setminus F$,
  then each of $\hat \Gamma_{x_i / x_j = a}$, $\hat \Gamma_{x_i / x_j \approx
    a^-}$, and $\hat \Gamma_{x_i / x_j \approx a^+}$ includes $\Gamma_{x_i
    / x_j \approx a}$. Thus every set $\hat \Gamma \in \hat {\mathcal S}$
  includes a set $\Gamma \in \mathcal S$. So, if the consistency
  criterion is satisfied by some $\hat \Gamma \in \hat {\mathcal S}$, it is
  satisfied by some $\Gamma \in {\mathcal S}$.\qed

We now parameterize the finite subsets of each $\hat \Gamma \in
\hat{\mathcal S}$. For each $\varepsilon > 0$, we define a formula
\[
\hat \Gamma[\varepsilon] = \bigwedge_{i < j} \hat \Gamma_{i, j}[\varepsilon],
\]
where
\begin{enumerate}
\item $\hat \Gamma_{x_i / x_j = a}[\varepsilon]$ is the formula $x_i = a x_j$
\item $\hat \Gamma_{x_i / x_j \approx r}[\varepsilon]$ is $(r -
  \varepsilon) x_j < x_i < (r + \varepsilon) x_j$
\item $\hat \Gamma_{x_i /x_j \approx \infty}$ is $x_i > (1 / \varepsilon) x_j$
\item $\hat \Gamma_{x_i / x_j \approx a^-}$ is $(a - \varepsilon)
  x_j < x_i < a x_j$
\item $\hat \Gamma_{x_i / x_j \approx a^+}$ is $a x_j < x_i < (a + \varepsilon)
  x_j$
\end{enumerate}
For every $\varepsilon$, $\hat \Gamma[\varepsilon]$ is implied by some
finite subset of $\hat \Gamma$. Conversely, every finite subset of
$\hat \Gamma$ is implied by $\hat \Gamma[\varepsilon]$ for some
$\varepsilon > 0$, and, in fact, for an $\varepsilon$ of the form $1
/n$ for some $n \in \NN$. Thus the consistency criterion for $\lnot
\ph$ is equivalent to the following:
\begin{quote}
  there is a set $\hat \Gamma \in \hat {\mathcal S}$ such that for
  every $\varepsilon > 0$, the sentence
\[
\ex {\vec x} (\hat \Gamma[\varepsilon] \land \lnot \phadd(\vec x)) \land
\ex {\vec x} (\hat \Gamma[\varepsilon] \land \lnot \phmult(\vec x))
\]
is true of the real numbers. 
\end{quote}
The sets $\hat \Gamma \in \hat {\mathcal S}$, and the corresponding
formulas $\hat \Gamma[\varepsilon]$, are parameterized by tuples of symbols
from the set
\[
\{ \mbox{`$\mathord{=}a$'} \st a \in \RR \} \cup 
\{ \mbox{`$\mathord{\approx} r$'} \st r \in \RR \setminus F \} \cup 
\{ \mbox{`$\infty$'} \} \cup 
\{ \mbox{`$\mathord{\approx} a^-$'} \st a \in \RR \} \cup 
\{ \mbox{`$\mathord{\approx} a^+$'} \st a \in \RR \}.
\]
When $F = \RR$, there are no sets with parameters of the second kind,
and so the consistency criterion can be expressed in the language of
real closed fields. By Theorem~\ref{alg:maximal:thm}, $T[\RR]$ is a
conservative extension of $T[\AAA]$. Thus we have:
\begin{theorem}
The universal fragment of $T[\AAA]$ is decidable.\qed
\end{theorem}

When $F$ is a proper subfield of $\RR$, the revised consistency
criterion for $\lnot \ph$ can be expressed as a sentence of the form
\[
\ex {\vec r \in \RR \setminus F} \ex {\vec a \in \RR} \fa {\varepsilon
  > 0} \ex {\vec x, \vec x'} \theta
\]
where $\theta$ is a quantifier-free formula in the language of real
closed fields. By quantifier-elimination for real closed fields, this
is equivalent to a sentence of the form $\ex{\vec r \in \RR \setminus
  F} \eta$, where $\eta$ is a quantifier-free formula in the language
of real closed fields. Say $F$ is a \emph{sufficiently computable}
subfield of $\RR$ if $F$ is a computable subfield of $\RR$ and there
is an algorithm to determine whether a real algebraic number $a$
(described in terms of a definition, say, in the language of real
closed fields) is in $F$.

\begin{theorem}
\label{main:decidability:thm}
For any sufficiently computable $F \subseteq \RR$, the universal fragment
of $T[F]$ is decidable.
\end{theorem}

By our analysis of the consistency criterion,
Theorem~\ref{main:decidability:thm} is a consequence of the following:

\begin{theorem}
\label{definability:theorem}
For any sufficiently computable $F \subseteq \RR$, there is an algorithm
to decide whether a sentence of the form $\ex {\vec x \in \RR
  \setminus F} \ph(\vec x)$ holds of the reals, where $\ph$ is a
formula in the language of real closed fields.
\end{theorem}

We will prove something more general. Let $R$ be any real closed
field. A function $h(\vec x)$ or a predicate $E(\vec x)$ on $R$ is
said to be \emph{semialgebraic} if it is definable in the language of
real-closed fields without parameters.

\begin{theorem}
\label{definability:theorem:two}
Let $R$ be any real closed field, and let $F$ be any proper subfield
of $R$. If $E,h_1, \ldots, h_m$ are semialgebraic, then
\[
\ex {x_1 \not\in F} \ldots \ex {x_n \not\in F} (E(\vec x, \vec y)
\land h_1(\vec x, \vec y) \not\in F \land \ldots \land h_m(\vec x,
\vec y) \not\in F)
\]
is equivalent to a positive boolean combination of assertions of the
form $D(\vec y)$ and $g(\vec y) \not\in F$, where $D$ and $g$ are
semialgebraic.  Furthermore, there is an algorithm for determining an
expression of this form from (presentations of) $E, h_1, \ldots, h_m$.
This algorithm does not depend on $R$ or $F$.
\end{theorem}

In particular, when there are no variables $\vec y$,
Theorem~\ref{definability:theorem:two} asserts that any assertion of
the form $\ex {\vec x \in \RR \setminus F} E(\vec x)$ is effectively
equivalent to a boolean combination of sentences in the language of
real-closed fields and assertions of the form $g \not\in F$, where $g$
is a real algebraic constant. Thus
Theorem~\ref{definability:theorem:two} implies
Theorem~\ref{definability:theorem}.

\proof
  We use induction on $n$. When $n = 0$ there is nothing
  to do. Suppose the theorem is true for $n$. Then
\[
\ex {x_1 \not\in F} \ldots \ex {x_{n+1} \not\in F} (E(\vec x, \vec y)
\land h_1(\vec y) \not\in F \land \ldots \land h_m(\vec y) \not\in F)
\]
is equivalent to $\ex {x_1 \not\in F} \psi(x_1, \vec y)$, where $\psi$
has the requisite form. We can then write $\psi$ as a disjunction of
formulas of the form
\[
D(x_1, \vec y) \land g_1(x_1,
\vec y) \not \in F \land \ldots \land g_l(x_1, \vec
y) \not \in F
\]
where $D, g_1, \ldots, g_l$ are semialgebraic.  Since we can factor
the existential quantifier $\exists {x_1}$ across the disjunction, it
suffices to prove Theorem~\ref{definability:theorem:two} for the
special case $n = 1$.

So, resorting to the original notation, let $E(x,\vec y), h_1(x, \vec
y), \ldots, h_m(x, \vec y)$ be semialgebraic. We need to show that
\begin{equation}
\label{definability:eqn}
\ex {x \in \RR \setminus F} (E(x,\vec y) \land h_1(x,\vec y) \not\in F \land \ldots
h_m(x,\vec y) \not\in F)
\end{equation}
is equivalent to a positive boolean combination of assertions $D(\vec
y)$ and $g(\vec y) \not\in F$, for semialgebraic $D$ and $g$.

By the theory of definability in real closed fields
\cite{basu:et:al:03,van:den:dries:98}, for each fixed $\vec y$, the
set $\{x \st E(x,\vec y) \}$ is a finite union of disjoint intervals
(including intervals of the form $(-\infty, a)$, $(-\infty, a]$, $(a,
\infty)$, and $[a, \infty)$) with endpoints that are definable in the
parameters $\vec y$. Similarly, fixing $\vec y$, for all but finitely
many points $x$ of $R$ all the functions $h_i$ are either locally
increasing or locally decreasing or locally constant at $x$. A bound
$p$ on the number of such intervals and exceptional points,
independent of $\vec y$, can be determined effectively from the
presentations of $E, h_1, \ldots, h_m$. Furthermore, for fixed $n$,
terms like ``the left endpoint of the $n$th interval (in increasing
order) in the decomposition of $\{ x \st E(x,\vec y)\}$, if there is
one, or $0$ otherwise'' and ``the $n$th point at which one of the
$h_i$'s is neither locally monotone nor locally constant, if there is
one, or $0$ otherwise'' are semialgebraic functions of $\vec y$.

As a result, for each fixed $\vec y$, there is a sequence of at most
$p$ disjoint nonempty open intervals $J_1, \ldots, J_q$ and at most $p$
exceptional points $u_1, \ldots, u_r$ such that
\begin{itemize}
\item $\{ x \st E(x, \vec y) \} = J_1 \cup \ldots \cup J_q \cup \{
  u_1, \ldots, u_r \}$, and
\item on each interval $J_n$, all the functions $h_i$ are either
  monotone or constant.
\end{itemize}
Furthermore, all the following are semialgebraic in $\vec y$:
\begin{itemize}
\item the predicates $D_{q,r}(\vec y)$, where $q,r \leq p$, which
  assert that there are exactly $q$ intervals in the decomposition of
  $\{ x \st E(x,\vec y) \}$ and $r$ exceptional
  points;
\item the predicate $G_{i,n}(\vec y)$ which asserts that $h_i$ (as
  a function of $x$), is constant on
  $J_n$; and
\item the functions $k_{i,n}(\vec y)$ which return the value of $h_i$
  on $J_n$, if $h_i$ is constant on $J_n$, or $0$ otherwise.
\end{itemize}

Given $\vec y$, assuming that there are $q$ intervals $J_n$ and $r$
exceptional points, we claim that (\ref{definability:eqn}) is
equivalent to the following disjunction:
\begin{enumerate}
\item there is an interval $J_n$, $n = 1, \ldots, q$, such that for
  each function $h_i$, if $h_i$ is constant on $J_n$, then the value
  of $h_i$ on $J_n$ is not in $F$; or
\item for one of the exceptional points $u_n$, $n = 1, \ldots, r$, we
  have $u_n \not \in F,$ and $h_i(u_n) \not \in F$ for each $i$.
\end{enumerate}
By the preceding paragraph, this can be expressed as a positive
boolean combination $\psi_{q,r}(\vec y)$ of assertions of the form
$H(\vec y)$ and $l(\vec y) \not\in F$, where $H$ and $l$ are
semialgebraic. This means that the expression
\[
\bigvee_{q, r \leq p} (D_{q,r}(\vec y) \land \psi_{q,r}(\vec
y))
\]
is of the requisite form. Thus, to complete the proof of
Theorem~\ref{definability:theorem:two}, it suffices to establish the
equivalence of (\ref{definability:eqn}) with the disjunction of 1 and 2.

Suppose (\ref{definability:eqn}) holds, and, given $\vec y$, let $x
\not\in F$ witness the existential quantifier. Since $E(x,\vec y)$
holds, either $x$ is in $J_n$ for some $n$, in which case clause 1
holds, or $x$ is one of the exceptional points $u_n$, in which case
clause 2 holds.

Conversely, given $\vec y$, suppose either 1 or 2 holds. If 2
holds, then that exceptional value $u_n$ witnesses the existential
quantifier in (\ref{definability:eqn}). So assume 1 holds, and let
$J$ be an interval on which all the functions that are constant take a
value not in $F$. Renumbering, let $h_1, \ldots, h_l$ be functions
that are not constant on $J$. It suffices to show that there is an $x
\in J \setminus F$ such that $h_1(x,\vec y), \ldots, h_l(x,\vec y)$
are not in $F$.

We consider two cases. First, suppose $R$ properly contains the real
algebraic closure of $F(\vec y)$ in $R$. Then one can choose an $x$
transcendental over $F(\vec y)$ in the interval $J$. This $x$ has the
desired property: if $h_i(x,\vec y) = a$ for some $i = 1\ldots l$,
then $h_i(x,\vec y) - a = 0$ is a nontrivial algebraic identity in
$\vec y$ and elements of $F$, contradiction. Otherwise, $R$ is equal
to the real algebraic closure of $F(\vec y)$ in $R$. Since $F$ is
properly contained in $R$, we can choose an $x$ with sufficiently high
algebraic degree over $F(\vec y)$, in which case an equality
$h_i(x,\vec y) = a$ for some $i = 1\ldots l$ again yields a
contradiction.\qed

Note that in the instance of Theorem~\ref{definability:theorem:two}
needed for Theorem~\ref{definability:theorem}, $R = \RR$ and $F$ is a
countable subfield, in which case the implication from 1 to
(\ref{definability:eqn}) in the last paragraph of the preceding proof
follows more easily from cardinality considerations.


\section{Normal forms}
\label{normal:forms:section}

When dealing with an associative and commutative operation like
addition, it is common to put terms in an appropriate normal form. For
example, one can always rearrange a sum $t_1 + \ldots + t_n$ so that
parentheses are associated, say, to the left, and $t_1, \ldots, t_n$
are ordered according to a fixed ordering of terms; this makes it easy
to tell whether or not two such sums agree up to the associativity and
commutativity of addition. In the theories $T[F]$, not only do we have
addition and multiplication (as well as subtraction and division), but
also multiplication by constants from $F$. In this section, we will
show that one can still, fruitfully, put terms in $T[F]$ into a normal
form. This provides an algorithm for testing whether two terms are
provably equal: just put them in normal form, and compare. 

In fact, to show that normal forms are unique, we will take care to
define an ordering on these terms that is compatible with the axioms
for $<$ in $T[F]$. This will enable us to construct a term model of
$T[F]$ in which different terms in normal form denote different
elements. It will also enable us to show that any equality between
terms that can be established in $T[F]$ can be proved without using the
ordering. 

We define a set of \emph{preterms} inductively, each with an
associated rank, as follows. For each $n$, a preterm of rank $2n + 1$
is called an ``additive preterm,'' and a preterm of rank $2n + 2$
is called a ``multiplicative preterm.'' A preterm of rank $0$ is
called a ``basic preterm.''
\begin{itemize}
\item Each variable, $x, y, z, \ldots$ is a preterm of rank $0$, as
  well as the constant, $1$.
\item For $n$ greater than $0$ and odd, if $t_1, \ldots, t_k$ are
  multiplicative or basic preterms of rank at most $n-1$, $k \geq 2$,
  $a_1, \ldots a_k$ are nonzero elements of $F$, and at least one
  $t_i$ has rank $n-1$, then $a_1 t_1 + a_2 t_2 + \ldots + a_k t_k$ is
  a preterm of rank $n$.
\item For $n$ greater than $0$ and even, if $t_1, \ldots, t_k$ are
  additive or basic preterms of rank at most $n-1$ and other than $1$,
  $i_1, \ldots, i_k$ are nonzero integers, either $k \geq 2$ or $i_1
  \neq 1$, and at least one $t_i$ has rank at least $n - 2$, then $t_1^{i_1}
  t_2^{i_2} \cdots t_k^{i_k}$ is a preterm of rank $n$.
\end{itemize}
Here parentheses in products and sums are assumed to associate to the
left, and for an integer $i$, $t^i$ is the $i$-fold product of $t$
with itself if $i$ is positive, or 1 divided by the $-i$-fold product
of $t$ with itself if $i$ is negative.  Note that there is no constant
multiplier for multiplicative preterms. The condition ``$k \geq 2$ or
$i_1 \neq 1$'' in the third clause allows $x^2$, for example, but
rules out $x^1$.

We now define, simultaneously, a normal form for preterms together
with an ordering $s \prec t$ on preterms in normal form. We assume
that variables have been indexed $x_1, x_2, \ldots$. For each
$n$, we define the notion of normal form, as well as the ordering, for
terms of rank at most $n$, as follows:
\begin{enumerate}
\item $n = 0$: Each basic preterm is in normal form. These are ordered
  $1 \succ x_1 \succ x_2 \succ \ldots$
\item $n > 0$, odd: An additive preterm $a_1 t_1 + a_2 t_2 + \ldots +
  a_k t_k$ is in normal form if and only if each $t_i$ is in normal
  form, $t_1 \succ t_2 \succ \ldots \succ t_k$, and $a_1 = 1$.
  
  To define $s \prec t$ when at least one of $s$ and $t$ has rank $n$
  and the other has rank at most $n$, write
\[
s = a_1 u_1 + a_2 u_2 + \ldots + a_k u_k
\]
and
\[
t = b_1 u_1 + b_2 u_2 + \ldots + b_k u_k
\]
where $u_1 \succ u_2 \succ \ldots \succ u_k$ are preterms of rank at most
$n-1$, and now the $a_i$'s and $b_i$'s are allowed to be $0$. Then use
lexicographic order: $s \prec t$ if and only if $a_i \neq b_i$ for
some $i$ and $a_i < b_i$ for the least such $i$.

\item $n > 0$, even: A multiplicative preterm $t_1^{i_1} t_2^{i_2}
  \cdots t_k^{i_k}$ is in normal form if and only if each $t_m$ is in
  normal form, and $t_1 \succ t_2 \succ \ldots \succ t_k$. To compare
  two multiplicative preterms of rank at most $n$, the procedure is
  slightly more complicated now, since we now consider the standing of
  the subterms in relation to the basic preterm $1$. Write the subterms
  $s_i$ occurring in $s$ and the subterms $t_j$ occurring $t$,
  together with the preterm 1, in $\succ$-decreasing order as $u_1,
  \ldots, u_m, 1, u_{m+1}, \ldots, u_k$. Then express
\begin{equation}
\label{normal:form:one:eq}
s = u_1^{i_1} u_2^{i_2} \cdots u_m^{i_m} \cdot 1 \cdot u_{m+1}^{i_{m+1}}
  \cdots u_k^{i_k}
\end{equation}
and
\begin{equation}
\label{normal:form:two:eq}
t = u_1^{j_1} u_2^{j_2} \cdots u_m^{j_m} \cdot 1 \cdot u_{m+1}^{j_{m+1}}
  \cdots u_k^{j_k}
\end{equation}
where now the $i_n$'s and $j_n$'s may be $0$. We now say $s \prec t$
  if and only if
\begin{itemize}
\item there is an $n \leq m$ such that $i_n \neq j_n$, and, for the
  least such $n$, $i_n < j_n$; or
\item For every $n \leq m$, $i_n = j_n$, but there is some $n > m$
  such that $i_n \neq j_n$, and $i_n > j_n$ for the \emph{largest}
  such $n$. 
\end{itemize}
\end{enumerate}

Note that the clause 1 of the definition of $\succ$ makes sense if we
think of the variables as being positive values, with each $x_{i+1}$
infinitesimally small compared to $x_i$ and 1. Clause 2, which treats
the case where the term of highest rank is additive, is also
intuitively consistent with an interpretation of $\prec$ as denoting a
relation, ``is infinitely smaller than,'' on positive numbers. Clause
3, which treats the case where the term of highest rank is
multiplicative, has similarly been designed to admit such an
interpretation. The main constraint there was to ensure that the
ordering cohere, in the following sense:

\begin{lemma}
\label{normal:form:lemma}
Let $n > 0$ be even, and let $s$ and $t$ be preterms of rank less than
or equal to $n$. Then the ordering of $s$ and $t$ is equivalent to the
order obtained under clause 3, when $s$ and $t$ are put in the form
(\ref{normal:form:one:eq}) and (\ref{normal:form:two:eq}),
respectively.\qed
\end{lemma}

Lemma~\ref{normal:form:lemma} is needed to prove Lemma~\ref{nf:lemma}.
The proof proceeds by running through the cases where each of $s$ and
$t$ is a variable, the constant 1, an additive term, or a
multiplicative term. For example, if $s$ and $t$ are additive and $1
\succ s \succ t$, one easily verifies that $1 s^1 t^0 \succ 1 s^0 t^1$
under Clause 3. The other cases are similarly straightforward.

Say that a term is in \emph{normal form} if it is either $0$ or of the
form $a t$, where $t$ is a preterm in normal form and $a$ is a nonzero
element of $F$. Let $T'_\fn{add}[F]$ be the restriction of $\Tadd[F]$
to the language without the ordering $<$. Let $T'_\fn{mult}[F]$ be
corresponding restriction of $\Tmult[F]$. Let $T'[F] = T'_\fn{add}[F]
\cup T'_\fn{mult}[F]$. It is straightforward to verify the following:

\begin{theorem}
\label{nf:theorem:a}
For every term $t$, there is a term $\hat t$ in normal form, such that
$T'[F]$ proves $t = \hat t$.\qed
\end{theorem}

Our main goal, in this section, is to prove the following:
\begin{theorem}
\label{nf:theorem}
If $\hat s$ and $\hat t$ are terms in normal form, and $T[F]$ proves
$\hat s = \hat t$, then $\hat s = \hat t$.
\end{theorem}
Note that the last equality is \emph{syntactic} equality; in other
words, $T$ proves that two terms in normal form are equal if and only
if they are the same term. 

As corollaries, we obtain the following:

\begin{corollary}
  There is an efficient procedure for determining whether $T[F]$
  proves $s = t$.
\end{corollary}

\proof
Just put $s$ and $t$ in normal form, and compare.\qed

\begin{corollary}
$T[F]$ and $T'[F]$ have the same provable equalities.
\end{corollary}

\proof
  If $T[F]$ proves $s = t$, then $s$ and $t$ have the same normal form
  $u$. Since $T'[F]$ proves $s = u$ and $t = u$, it proves $s = t$.\qed

To prove Theorem~\ref{nf:theorem}, first let us extend the ordering
$\prec$ from preterms in normal form to terms in normal form, as
follows: if $s$ and $t$ are preterms in normal form, then
\begin{itemize}
\item $0 \prec a t$ if and only if $a > 0$.
\item $a t \prec 0$ if and only if $a < 0$.
\item $0 \not \prec 0$
\item $a s \prec b t$ if and only if:
\begin{itemize}
\item $a$ is negative, and $b$ is positive
\item $a$ and $b$ are both positive, and either $s \prec t$ or $s = t$
  and $a < b$
\item $a$ and $b$ are both negative, and either $s \succ t$ or $s = t$
  and $a < b$
\end{itemize}
\end{itemize}
It suffices to show
\begin{lemma}
\label{nf:lemma}
  There is a model $\mdl M$ of $T[F]$ such that if $\hat s$ and $\hat
  t$ are terms in normal form and $\hat s \prec \hat t$, then $\hat
  s < \hat t$ holds in $\mdl M$.
\end{lemma}

\proof
  Note that operations of addition, subtraction, multiplication, and
  division are naturally defined on terms in normal form. For example,
  suppose $a (a_1 s_1 + a_2 s_2 + \ldots + a_k s_k)$ and $b (b_1 t_1 +
  b_2 t_2 + \ldots + b_l s_l)$. To express their sum as a term in
  normal form, multiply through by $a$ and $b$, respectively, combine
  terms, and express the sum as $c_1 u_1 + c_2 u_2 + \ldots + c_m
  u_m$, where $u_1 \succ u_2 \succ \ldots \succ u_m$ and each $c_i
  \neq 0$, or $0$. In the former case, the desired normal-form term is
  $c_1 (u_1 + (c_2 / c_1) u_2 + \ldots + (c_m / c_1) u_m)$. This term
  model \emph{almost} satisfies the claim of Lemma~\ref{nf:lemma};
  it satisfies all the axioms of $T[F]$ indicated in
  Section~\ref{theories:section}, except for the axiom that asserts
  that the multiplicative group of positive elements is divisible.
  That is, all that is missing are $n$th roots of positive elements.
  To remedy the situation, we embed this term model in an
  expanded set of formal terms, defined as follows.

  Let $F'$ be the smallest subfield of $\RR$ that includes $F$ and is
  closed under $n$th roots of positive elements, for positive $n$.
  Define the set of \emph{extended preterms} inductively, as above,
  with the following changes:
\begin{itemize}
\item in the additive extended preterms $a_1 t_1 + \ldots + a_k t_k$,
  the coefficients $a_i$ are taken from $F'$; and
\item multiplicative extended preterms are taken to be formal products
  $t_1^{i_1} t_2^{i_2} \ldots t_k^{i_k}$ where now the exponents $i_j$
  are \emph{rational numbers}.
\end{itemize}
Define the set of extended preterms in normal form, the ordering on
these, the set of extended terms in normal form, and the ordering on
these, exactly as before. Once again, operations of addition and
multiplication can be defined on extended terms in normal form.
Lemma~\ref{normal:form:lemma}, as well as the analogue for additive
preterms, carry over to extended preterms as well.

Let $\mdl M$ be the model whose universe is the set of extended terms
in normal form, with the associated ordering and operations of
addition and multiplication. Clearly there is an embedding of the
set of terms in normal form into the set of extended terms in normal
form which preserves all the operations. So it suffices to show that
$\mdl M$ satisfies $T[F]$.

We simply run through the axioms given in
Section~\ref{theories:section}. Verifying the axioms of
$\Tcommon[F]$ is straightforward, as well as the fact that the terms
form an abelian group under addition, and the positive terms form an
abelian group under multiplication.

To show that the ordering is compatible with multiplication of
positive elements, we need to show that $s \prec t \limplies s u \prec
t u$ holds of positive terms $s, t, u$ in normal form. Let $s = a s'$,
$t = b t'$, and $u = cu'$ where $s'$, $t'$, and $u'$ are preterms in
normal form, and $a$, $b$, and $c$ are positive. Then $s u = (ac)
s'u'$ and $tu = (b c)t'u'$. Since $s \prec t$, we have either $s'
\prec t'$, or $s' = t'$ and $a < b$. In the first case,
Lemma~\ref{normal:form:lemma} and Clause 3 of the definition of $\prec
$ guarantees that $s' u' \prec t' u'$, and hence $s u \prec t u$. In
the second case, $s'u' = t' u'$ and $ac < bc$, so, again, $s u \prec t
u$.

Showing that the ordering is compatible with addition is similarly
straightforward. So we only need to show that the multiplicative group
of positive elements is divisible. Let $a t$ be an extended term in
normal form satisfying $at \succ 0$. Then $a > 0$, and we can view $t$
as a multiplicative preterm $t_1^{i_1} t_2^{i_1} \ldots t_k^{i_k}$,
possibly with $k = 1$ and $i_1 = 1$. But this has $n$th root
$\sqrt[n]{a} t_1^{i_1/n} t_2^{i_1/n} \ldots t_k^{i_k/n}$, where this
is identified with $\sqrt[n]{a} t_1$ if $k = 1$ and $i_1/n = 1$.\qed 

We note that the complicated definition of $\prec$ in the
multiplicative clause of the ordering of preterms was designed to
ensure that $\prec$ is compatible with the axioms of $T[F]$. This, in
turn, was used to construct the term model in the proof of
Theorem~\ref{nf:theorem}. Theorem~\ref{nf:theorem} remains true,
however, for a simpler version of $\prec$, in which we simply use a
lexicographic ordering at the multiplicative stage. This simpler
ordering, and the associated normal forms, are more amenable to
implementation. (Indeed, it may also be natural to order terms of
lower rank before terms of higher rank.) To derive the variant of
Theorem~\ref{nf:theorem} for these normal forms, it suffices to show
that the map from terms in the simpler normal form to the normal form
we have used here is injective.  In other words, it suffices to show
that if $s$ and $t$ are in the simpler normal form, $u$ a term in the
normal form we have used here, and $T[F]$ proves both $s = u$ and
$t=u$, then $s$ and $t$ are syntactically identical. This can be done
by a careful induction on the maximum rank of $s$ and $t$.

Note also that it is harmless, and again useful from an implementation
point of view, to extend the language of $T[F]$ to include
exponentiation to arbitrary integers. Since $n$th roots of positive
elements can be defined in $T[F]$, one can similarly expand the
language of $T[F]$ to allow $n$th root functions for positive $n$, or
even exponentiation to any rational power. One has to be careful,
however, to provide a consistent interpretation of the $n$th root
function on negative elements, and natural simplifications may depend
on knowing the sign of the relevant terms. For example, $\sqrt{x^2}$
can be simplified to $x$ if $x$ is positive and $-x$ if $x$ is
negative. For that reason, determining an appropriate normal form
representation for terms involving $n$th roots is more complicated.
Similar complications arise in obtaining an adequate handling of
absolute value, max, and min. The issue of obtaining useful canonical
representations for such extensions is of practical importance, and is
discussed further in Section~\ref{extensions:section} below.

Finally, we note that the method of computing normal forms only gives
a decision procedure for provable equations in the absence of
hypotheses. For example, $T[F]$ proves $1 + x^2 + y^2 \neq 0$ (or,
equivalently, $1 + x^2 + y^2 = 0 \limplies 0 = 1$), but this is not
provable in $T'[F]$.


\section{Building models of  $T[F]$}
\label{models:section}

In Sections~\ref{undecidability:one:section} and
\ref{undecidability:two:section}, our goal will be to prove
undecidability results (and conditional undecidability results) for the
theories $T[F]$. Recall the alternative formulations $T[F]^*$
introduced in Section~\ref{theories:section}, in the language with
symbols $0, 1, +, \times, <$ and constants $c_a$ for each $a \in F$.
In light of Theorem~\ref{und:one:three}, we will work exclusively with
the theories $T[F]^*$. Our strategy will be to build models of
$T[F]^*$ in which $F$ and $\ZZ$ are, respectively, definable. In this
section, we will develop techniques for building such models.

Let $\RRR = \la R,<,+,-,\times \ra$ be an ordered real closed field
extending the countable ordered subfield $F \subseteq \RR$. More
specifically, we assume that $F$ is a subfield of $\RRR$, where the
ordering on $F$ agrees with the ordering in $\RRR$.

\begin{definition}
We say that $h$ is an \emph{$F$-bijection} of $\RRR$ if and only if
\begin{enumerate} 
\item $h:R \to R$ is an order preserving bijection.
\item $h(0) = 0$ and $h(1) = 1$.
\item For all $x \in R$ and $a \in F$, we have $h(ax) = ah(x)$.
\end{enumerate}
\end{definition}

Given an $F$-bijection $h$, we define the structure $h^{-1}[\RRR]$ in
the language of $T[F]^*$ as follows. The domain of $h^{-1}[\RRR]$ is
$R$. The symbols $0,1,+,$ and $<$ are interpreted as in $\RRR$. For $a
\in F$, $c_a$ is interpreted as $a$. The symbol $\times$ is
interpreted in $h^{-1}[\RRR]$ as $\timesstar$, defined by the equation
\[
x \timesstar y = h^{-1}(h(x)h(y)).
\]
It follows from the definition that $x \timesstar y = z$ if and only if
$h(x)h(y) = h(z)$. Hence $h$ is an isomorphism from $\la
R,\timesstar,<\ra$ onto $\la R,\times,<\ra$.

\begin{theorem}
  \label{und:two:one}
  Let $h$ be an $F$-bijection of $\RRR$. The model $h^{-1}[\RRR]$
  satisfies $T[F]^*$.
\end{theorem}

\proof
  Recall the axiomatization of $T[F]^*$ given in
  Section~\ref{theories:section}.  We first verify axioms 1,2 in
  $h^{-1}[\RRR]$. The group given by $0,+,<$ is obviously an ordered
  commutative group.  Since $h$ is an isomorphism from $\la
  R,\timesstar,< \ra$ onto $\la R,\times,< \ra$, we have that
  $1,\times,<$ is a divisible ordered commutative group on the
  positive elements of $R$.
  
  Axioms 3a-3c obviously hold in $h^{-1}[\RRR]$. For axioms 4a,4b,
  note that for all $a \in F$,
\[
a \timesstar x = h^{-1} (h(a)h(x)) = h^{-1}(ah(x)) = ah^{-1}(h(x)) =
ax.
\]
Hence 
\[
(a+b) \timesstar x = (a+b)x = ax + bx = a \timesstar x + b \timesstar x
\]
and
\[\hbox to 100 pt{\hfill}
a \timesstar (x+y) = a(x+y) = ax + ay = (a \timesstar x) + (a \timesstar y).
\hbox to 100 pt{\hfill}\qEd\]

So far, we have only assumed that $\RRR$ is an ordered real closed
field extending the countable ordered subfield $F \subseteq \RR$. We
will now need to assume that $\RRR$ obeys some additional conditions.
Note that $R$ is a densely ordered set. An \emph{interval} in $R$ is a
$J \subseteq R$ such that for all $x < y < z$, $x,z \in J$, $y \in R$,
we have $y \in J$.  $J$ is said to be \emph{nontrivial} if and only if $J$
has infinitely many elements. This is the same as saying that $J$ has
at least two elements.

By a standard saturation argument, we will fix an ordered real closed
field $\RRR$, such that the following hold:
\begin{enumerate}
\item $R$ is countable.
\item $\RRR$ extends $F$ in the sense above.
\item Let $n \geq 1$. Suppose that for all $i \geq 1$, $g_i,h_i:R^n
  \to R$ are $\RRR$-definable, where $n$ may depend on $i$. Then
  $\cup_i g_i[F^n]$ has an upper bound. Furthermore, suppose each
  $g_i[F^n]$ lies strictly below each $h_j[F^n]$. Then the interval
  strictly above each $g_i[F^n]$ and strictly below each $h_j[F^n]$ is
  nontrivial.
\end{enumerate}
Here, as always, $\RRR$-definability allows the use of parameters from
$R$, and the notation $f[S]$ denotes the forward image of $f$ on
$S$. The existence of such a field can be proved by starting with a
countable ordered real closed subfield $R_0$ of $\RR$ containing $F$,
and then building a countably infinite chain of elementary extensions.
At each stage, use compactness to ensure that the required upper
bounds in 3 exist, and also that there are $x < y$ forming the
required nontrivial intervals. (For similar constructions see, for
example, \cite[Chapter 5]{chang:keisler:90}.)
 
Below, we will refer to condition 3 as the ``saturation condition on
$F,\RRR$.'' We will use the terms ``lower bound'' and ``upper bound''
in the weak sense ($\le$, $\ge$), and we will use the terms ``strict
lower bound'' and ``strict upper bound'' in the strong sense ($<$,
$>$). For $x_1,\ldots,x_n \in R$, we write $F[x_1,\ldots,x_n]$ for the
subfield of $R$ obtained by adjoining $x_1,\ldots,x_n$ to $F$.

\begin{lemma}
\label{und:two:two}
Let $x_1,\ldots,x_n,y,z \in R$, where $y < z$. There exists $y < w < z$ such that $w$ is not algebraic over $F[x_1,\ldots,x_n]$.
\end{lemma}

\proof
  Let $x_1,\ldots,x_n,y,z$ be as given. Let $g_1,g_2,\ldots$ be
  $\RRR$-definable functions where the union of their images over
  appropriate Cartesian powers of $F$ consists of all elements
  $1/(u-y)$, where $u > y$ is algebraic over $F[x_1,\ldots,x_n]$. By
  the saturation property of $F,\RRR$, these elements have a strict
  upper bound $b$. Hence $y + 1/b$ is a strict lower bound on these
  elements. Set $w = y + 1/b$.\qed

Our goal in the next two sections will be to construct $F$-bijections
of $R$ such that properties of $\QQ$ or $\ZZ$ are coded into
$h^{-1}[\RRR]$. Our strategy will be to iteratively extend partial
$F$-homomorphisms until they become total and onto. The following
definitions and lemmas will support our constructions.

\begin{definition}
Let $V[F,\RRR]$ be the family of all sets $E \subseteq R$ such that
for some $x_1,\ldots,x_n \in R$, $n \geq 0$,
\[
E = \{ax_i \st 1 \le i \le n \land  a \in F\}.
\]
Let $W[F,\RRR]$ be the set of all partial one-one functions h from $R$
into $R$ such that the following hold:
\begin{enumerate}
\item $\dom(h) \in V[F,\RRR]$.
\item $h$ is order preserving.
\item $h(0) = 0$ and $h(1) = 1$.
\item For all $x \in \dom(h)$ and $a \in F$, we have $h(ax) = ah(x)$.
\end{enumerate}
\end{definition}

Note that for all $h \in W[F,\RRR]$, $\rng(h) \in V[F,\RRR]$. 

\begin{lemma}
\label{und:three:one}
Every $E \in V[F,R]$ is the image of an $\RRR$-definable function on
some $F^n$. Every $h \in W[F,R]$ is the restriction of an
$\RRR$-definable function to its domain.
\end{lemma}

\proof
  The first claim follows immediately from the definition. For the
  second claim, fix $x_1,\ldots,x_n \in R$ such that $\dom(h) = \{ax_i
  \st 1 \le i \le n a \in F\}$. Then $h = h_1 \cup \ldots \cup h_n$,
  where each $h_i:\{ax_i \st a \in F\} \to \{ah(x_i): a \in F\}$ is
  given by $h_i(ax_i) = ah_i(x_i)$.\qed

\begin{lemma}
\label{und:three:two}
For all $h \in W[F,\RRR]$, $h^{-1} \in W[F,\RRR]$.  
\end{lemma}

\proof
  Let $h \in W[F,\RRR]$. For all $x,y \in \rng(h) = \dom(h^{-1})$, if
  $x < y$, then $h(h^{-1}(x)) < h(h^{-1}(y))$, and so $h^{-1}(x) <
  h^{-1}(y)$. Similarly, $h^{-1}(0) = h^{-1}(h(0)) = 0$ and $h^{-1}(1)
  = h^{-1}(h(1)) = 1$. For any $a$ in $F$ and $x$ in $\rng(h)$,
  $h^{-1}(ax) = h^{-1}(h(a) h(h^{-1}(x))) = h^{-1}(h(a h^{-1}(x))) = a
  h^{-1}(x)$, as required.\qed

The following proposition provides a connection between types over
$\Tcommon[F]$, which were discussed in
Section~\ref{provability:section}, and the elements of $W[F,\RRR]$.

\begin{proposition}
\label{temp:prop}
Let $x_1, \ldots, x_n, y_1, \ldots, y_n$ be elements of $R$. Then
there is an $h \in W[F,\RRR]$ satisfying $h(x_i) = y_i$ for every $i$
if and only if $\vec x$ and $\vec y$ have the same types over
$\Tcommon[F]$.
\end{proposition}

We will not use Proposition~\ref{temp:prop} below, and so we omit the
proof, which is straightforward.

We now determine ways in which elements of $W[F,\RRR]$ can be
extended. We write $\fld(h)$ for $\dom(h) \cup \rng(h)$. The
\emph{$F$-multiples} of $x \in R$ are the elements $ax$, for $a \in
F$. We write $h \subseteq_1 h'$ if and only if the following hold:
\begin{enumerate}
\item $h,h' \in W[F,\RRR]$.
\item $h \subseteq h'$.
\item There exists $x \in \dom(h')\setminus \dom(h)$ such that
  $\dom(h') = \dom(h) \uplus \{ax \st a \in F \setminus \{ 0 \} \}$.
\end{enumerate}
Here, $\uplus$ denotes a disjoint union. Then $h \subseteq_1 h'$ is
equivalent to the following assertions.
\begin{enumerate}
\item $h,h' \in W[F,\RRR]$.
\item $h \subseteq h'$.
\item[3'] There exists $y \in \rng(h')\setminus \rng(h)$ such that
  $\rng(h') = \rng(h) \uplus \{ay \st a \in F \setminus \{ 0 \} \}$.
\end{enumerate}
Note that $h \subseteq_1 h'$ if and only if $h^{-1} \subseteq_1
h'^{-1}$. Note also that in 3,3' above, $x$ and $y$ are not unique, but
they are unique up to multiplication by an element of $F$.

\begin{lemma}
\label{und:three:three}
Let $h \in W[F,\RRR]$ and $x \in R \setminus \dom(h)$, $x > 0$. There
exists a nontrivial interval $J$ such that the following holds: for
all $y \in J$, there exists $h \subseteq_1 h'$ such that $h'(x) = y$.
\end{lemma}

\proof
  Let $h,x$ be as given. Obviously $\rng(h) = h[\dom(h)\res_{<x}]
    \uplus h[\dom(h)\res_{>x}]$, where $h[\dom(h)\res_{<x}]$ lies
    strictly below $h[\dom(h)\res_{>x}]$.
    
    \emph{Case 1.} $\dom(h)\res{>x}$ is empty. Let $J$ be the interval
    of elements of $R$ strictly above $\rng(h)$. By
    Lemma~\ref{und:three:one} and the saturation property of $F,\RRR$,
    $\fld(h)$ has a strict upper bound. Hence $J$ is nontrivial. Let
    $y \in J$, and define $h'(ax) = ay$, for all $a \in F$. We have
    only to verify that $h' \in W[F,\RRR]$.
    
    It suffices to show that $h'$ is order preserving. First, suppose
    $ax < a'x$, $a,a' \in F\setminus\{0\}$. Then $a < a'$, and so
    $h'(ax) = ah'(x) < a'h'(x) = h'(a'x)$.

    Next, suppose $v < ax$, $a \in F \setminus \{0\}$, $v \in \dom(h)$. If
    $a < 0$ then $v/-a > x$, which is impossible. Hence $a > 0$. Now
    $h(v/a) < h'(x)$. Hence $h(v) < ah'(x) = h'(ax)$.
    
    Finally, suppose $ax < v$, $a \in F\setminus \{0\}$, $v \in
    \dom(h)$. If $a > 0$ then $x < v/a$, which is impossible. Hence $a
    < 0$. Now $h(v/a) < h'(x)$, so $h(v)/a < h'(x)$, $h(v) >
    ah'(x) = h'(ax)$, and $h'(ax) < h(v)$.
    
    \emph{Case 2.} $\dom(h)\res_{<x}$ and $\dom(h)\res_{>x}$ are
    nonempty.  Let $J$ be the interval lying strictly above
    $h[\dom(h)\res_{<x}]$ and strictly below $h[\dom(h)\res_{>x}]$. By
    Lemma~\ref{und:three:one}, these two sets are each images of an
    $\RRR$-definable function on some $F^n$. Hence by the saturation
    condition on $F,\RRR$, $J$ is nontrivial. Let $y \in J$, and
    define $h'(ax) = ay$, for all $a \in F$. We have only to verify
    that $h' \in W[F]$.

    It suffices to show that $h'$ is order preserving. Suppose $ax <
    a'x$, $a,a' \in F\setminus \{0\}$. Then $a < a'$, and so $h'(ax) =
    ah'(x) < a'h'(x) = h'(a'x)$.

    Suppose $v < ax$, $a \in F\setminus \{0\}$, $v \in \dom(h)$. First
    assume $a > 0$. Then $v/a < x$, and so $h(v/a) < h'(x)$, $h(v)/a <
    h'(x)$, and $h(v) < ah'(x) = h'(ax)$. Now assume $a < 0$. Then
    $v/a > x$, and so $h(v/a) > h'(x)$, $h(v)/a > h'(x)$, and $h(v) <
    ah'(x) = a h'(ax)$.

    Finally, suppose $ax < v$, $a \in F\setminus \{0\}$, $v \in
    \dom(h)$. First assume $a > 0$. Then $x < v/a$, and so $h'(x) <
    h(v/a) = h(v)/a$, $ah'(x) < h(v)$, $h'(ax) < h(v)$. Now assume $a
    < 0$. Then $x > v/a$, and so $h'(x) > h(v/a) = h(v)/a$, $ah'(x) <
    h(v)$, and $h'(ax) < h(v)$.\qed

\begin{lemma}[First Extension Lemma]
\label{und:three:four} 
Let $h \in W[F]$ and $x \not\in \dom(h)$. Then there exists a
nontrivial interval $J$ such that the following holds: for all $y \in
J$, there exists $h \subseteq_1 h'$ such that $h'(x) = y$.
\end{lemma}

\proof
  Let $h$,$x$ be as given. The case $x > 0$ is given by
  Lemma~\ref{und:three:three}. So, suppose $x < 0$. Apply
  Lemma~\ref{und:three:three} to the case $-x > 0$, obtaining a
  nontrivial $J$ such that for all $y \in J$, there exists $h
  \subseteq_1 h'$ such that $h'(-x) = y$.

  We claim that $-J$ is a nontrivial interval such that for all $y \in
  -J$, there exists $h \subseteq_1 h'$ such that $h'(x) = y$. To see
  this, let $y \in -J$. Then $-y \in J$, and hence there exists $h
  \subseteq_1 h'$ such that $h'(-x) = -y$. But $h'(-x) = -y$ implies
  $h'(x) = y$, as required.\qed

\begin{lemma}[Second Extension Lemma]
\label{und:three:five}
Let $h \in W[F]$ and $x \not\in \rng(h)$. There exists a nontrivial
interval $J$ such that the following holds: for all $y \in J$, there
exists $h \subseteq_1 h'$ such that $h'(y) = x$.
\end{lemma}

\proof
  We obtain this from Lemma~\ref{und:three:four} as follows. Let
  $h,x$ be as given. Then $h^{-1} \in W[F]$ and $x \not\in
  \dom(h^{-1})$. By Lemma~\ref{und:three:three}, let $J$ be a
  nontrivial interval such that for all $y \in J$, there exists
  $h^{-1} \subseteq_1 h'$ such that $h'(x) = y$.

  We claim that for all $y \in J$, there exists $h \subseteq_1 h''$
  such that $h''(y) = x$. To see this, let $h \subseteq_1 h'$ be such
  that $h'(x) = y$. Then $h^{-1} \subseteq_1 h'^{-1}$ and $h'^{-1}(y)
  = x$.  That is, we can set $h'' = h'^{-1}$.\qed 


\section{Existential consequences of $T[F]$}
\label{undecidability:one:section}

The existential theory of $F$ consists of all sentences
\[
\ex {x_1,\ldots,x_n \in F} \ph(x_1,\ldots,x_n)
\]
where $\ph$ is a quantifier free formula involving $+$,$\times$,$<$,
and is interpreted in $\RR$. Here we show that the existential theory
of $F$ can be effectively reduced to the existential consequences of
$T[F]$ without auxiliary functions. This yields, in particular, a
conditional undecidability result for $T[\QQ]$; see
Corollary~\ref{und:four:six} below. 

We adhere strictly to the convention that if an equation holds, then
both sides must be defined. Also, a term is defined if and only if
each subterm is defined. For example,
\[
h^{-1}(h(x)h(1+x)) = x + h^{-1}(h(x)^2)
\]
implies that both sides of this equation are defined. In particular,
the above equation implies that $x,1+x \in \dom(h)$.

Let $h \in W[F,\RRR]$. We write $\alg(F,h)$ for the elements that are
algebraic over some $F[x_1,\ldots,x_n]$, $x_1,\ldots,x_n \in \fld(h)$.
We write $\trans(F,h)$ for $R\setminus \alg(F,h)$.

Note that by Lemma~\ref{und:three:one}, there exists
$x_1,\ldots,x_n \in R$ such that every element of $\alg(F,h)$ is
algebraic over $F[x_1,\ldots,x_n]$. This allows us to use
Lemma~\ref{und:two:two} to obtain an element of $\trans(F,h)$ in
every nontrivial interval.

\begin{lemma}
\label{und:four:one}
Let $h \in W[F,\RRR]$ be such that for every $x$, if
\[
h^{-1}(h(x)h(1+x)) = x + h^{-1}(h(x)^2),
\]
then $x \in F$. Let $b
\not\in \dom(h)$. There exists $h \subseteq_1 h' \in W[F,\RRR]$,
$h'(b)$ defined, such that $h'$ has the same property; i.e.~for every
$x$, if 
\[
h'^{-1}(h'(x)h'(1+x)) = x + h'^{-1}(h'(x)^2),
\]
then $x \in F$.
\end{lemma}

\proof
  Let $h$,$b$ be as given. By Lemmas~\ref{und:three:four} and
  \ref{und:two:two}, define $h \subseteq_1 h'$ such that $h'(b)
  \in \trans(F,h)$. We first show the conclusion for all $ab$, $a \in
  F$. We assume
\[
h'^{-1}(h'(ab)h'(1+ab)) = ab + h'^{-1}(h'(ab)^2).
\]
and derive a contradiction. Clearly $h'(ab)^2 = (ah'(b))^2 =
a^2h'(b)^2 \in \rng(h')$. Since $a \in F$ and $h'(b) \in \trans(F,h)$,
$a^2h'(b)^2 \in \rng(h')\setminus \rng(h)$, which consists of the
nonzero $F$-multiples of $h'(b)$. This contradicts that $h'(b) \in
\trans(F,h)$.

Finally, we show the conclusion for all $x \in \dom(h)\setminus F$. We
assume
\begin{equation}
\label{eq:und:one}
h'^{-1}(h'(x)h'(1+x)) = x + h'^{-1}(h(x)^2)
\end{equation}
and derive a contradiction. By the hypothesis on $h$,
(\ref{eq:und:one}) does not hold with $h'$ replaced by $h$. Hence if
we replace $h'$ by $h$, at least one side of (\ref{eq:und:one}) is
undefined.

\emph{Case 1.} $h(1+x)$ is undefined. Let $1+x = ab$, $a \in
F\setminus \{0\}$. Hence
\[
h'^{-1}(h(x)ah'(b)) = x + h'^{-1}(h'(ab^{-1})^2).
\]
Hence $h(x)h'(b) \in \rng(h')$. Since $h'(b) \in \trans(F,h)$,
$h(x)h'(b) \in \rng(h')\setminus \rng(h)$. Hence $h(x)h'(b)$ is a
nonzero $F$-multiple of $h'(b)$. This contradicts that $h(x) \not\in
F$.

\emph{Case 2}. $h(1+x)$ is defined, but $h^{-1}(h(x)h(1+x))$ is not
defined. Then $h(x)h(1+x)$ is a nonzero $F$-multiple of $h'(b)$.
Since $x \neq -1$, this product is nonzero. This contradicts that
$h'(b) \in \trans(F,h)$.

\emph{Case 3}. $h^{-1}(h(x)h(1+x))$ is defined, but $h^{-1}(h(x)^2)$
is undefined. Then $h(x)^2$ is a nonzero $F$-multiple of $h'(b)$. This
contradicts that $h'(b) \in \trans(F,h)$.\qed

\begin{lemma}
\label{und:four:two}
Let $h \in W[F,\RRR]$ be such that for every $x$, if
\[
h^{-1}(h(x)h(1+x)) = x + h^{-1}(h(x)^2)
\]
then $x \in F$. Let $b
\not\in \rng(h)$. Then there exists $h \subseteq_1 h' \in W$ such that
$h'^{-1}(b)$ is defined, and for every $x$, if 
\[
h'^{-1}(h'(x)h'(1+x))
= x + h'^{-1}(h'(x)^2)
\] 
then $x \in F$.
\end{lemma}

\proof
  Let $h$, $b$ be as given. By Lemmas~\ref{und:three:five} and
  \ref{und:two:two}, let $h \subseteq_1 h'$, where $h'^{-1}(b)
  \in trans(F,h)$. Write $c = h'^{-1}(b)$.

  We first show the conclusion for all $ac$, $a \in F \setminus \{ 0
  \}$. We assume
\[
h'^{-1}(h'(ac)h'(1+ac)) = ac + h'^{-1}(h'(ac)^2)
\]
and derive a contradiction. From the assumption, we have $1+ac \in
\dom(h')$. Since $c \in \trans(F,h)$, $1+ac \in \dom(h') \setminus
\dom(h)$. Hence $1+ac$ is a nonzero $F$-multiple of $c$. This
contradicts $c \in \trans(F,h)$.

Finally, we show the conclusion for all $x \in \dom(h)\setminus F$. We
assume
\begin{equation}
\label{eq:und:two}
h'^{-1}(h(x)h'(1+x)) = x + h'^{-1}(h(x)^2)
\end{equation}
and derive a contradiction. By the hypothesis on $h$,
(\ref{eq:und:two}) does not hold with $h'$ replaced by $h$. Hence if
we replace $h'$ by $h$, at least one side of (\ref{eq:und:two}) is
undefined.

\emph{Case 1.} $h^{-1}(h(x)^2)$ is undefined. Then $h(x)^2$ is a
nonzero $F$-multiple of $b$ and $h'^{-1}(h(x)^2)$ is a nonzero
$F$-multiple $ac$ of $c$. Clearly the left side of (\ref{eq:und:two})
either lies in $\dom(h)$ or is a nonzero $F$-multiple $ac$ of $c$.
Both possibilities contradict that $c \in \trans(F,h)$.

\emph{Case 2.}  $h^{-1}(h(x)^2)$ is defined and $h(1+x)$ is undefined.
Then $h(1+x)$ is a nonzero $F$-multiple of $b$ and $1+x$ is a nonzero
$F$-multiple of $c$. This contradicts that $c \in \trans(F,h)$.

\emph{Case 3.}  $h^{-1}(h(x)^2)$ and $h(1+x)$ are defined, but
$h^{-1}(h(x)h(1+x))$ is undefined. Hence $h(x)h(1+x)$ is a nonzero
$F$-multiple of $b$ and $h'^{-1}(h(x)h(1+x))$ is a nonzero
$F$-multiple of $c$. But the right side of (\ref{eq:und:two}) is
algebraic in $\fld(h)$. This is a contradiction.\qed

\begin{theorem}
\label{und:four:three}
There is a model $\mdl M$ of $T[F]^*$ with domain $R$, with the same
$0$,$1$,$+$,$<$ of $R$, in which for all $b$, $b(1+b) = b + b^2$
holds if and only if $b \in F$. In this equation, we use the
multiplication of $\mdl M$ to multiply $b$ and $1+b$.
\end{theorem}

\proof
  Let $h$ be the identity function on $F$. Then $h \in W[F,\RRR]$, and
  trivially we have that
\begin{itemize}
\item for every $x$, if $h^{-1}(h(x)h(1+x)) = x + h^{-1}(h(x)^2)$ then
  $x \in F$; and
\item for every $x \in F$, $h^{-1}(h(x)h(1+x)) = x + h^{-1}(h(x)^2)$.
\end{itemize}
Thus we can iterate Lemmas~\ref{und:four:one} and
\ref{und:four:two}, starting with the identity function on $F$,
diagonalizing over the countably many elements of $R$. We then obtain
$h \in W[F,\RRR]$ with domain $R$, such that 
\begin{quote}
for every $x$ in $R$, $h^{-1}(h(x)h(1+x)) = x + h^{-1}(h(x)^2)$ if and
only if $x \in F$.
\end{quote}
The required model $\mdl M$ of $T[F]^*$ is $h^{-1}[\RRR]$. Calculating
in $\mdl M$, we have
\[
x \timesstar (1+x) = h^{-1}(h(x)h(1+x))
\]
and
\[
x + (x \timesstar x) = x + h^{-1}(h(x)^2)
\]
Hence, for every $x$ in $R$, we have $x \timesstar (1+x) = x + (x
\timesstar x)$ if and only if $x \in F$, as required.\qed

\begin{corollary}
\label{und:four:four}
An existential sentence $\ph$ over $F$ in the language of ordered
fields is true if and only if in any model of $T[F]^*$, $\ph$ has
witnesses among the $b$ with $b(1+b) = b + b^2$.
\end{corollary}

\proof
  Suppose $\ph$ has the form $\ex{x_1, \ldots, x_n} \psi(x_1, \ldots,
  x_n)$ with $\psi$ quantifier-free, and suppose
  $\psi(a_1,\ldots,a_n)$ holds with $a_1,\ldots,a_n \in F$. Let $\mdl
  M$ be a model of $T[F]^*$. Then for all $1 \le i \le n$, $T[F]^*$
  proves $\ph(c_{a_1},\ldots,c_{a_n})$ and $c_{a_i}(1+c_{a_i}) =
  c_{a_i} + c_{a_i}^2$.

For the converse, Let $\mdl M$ be a model of $T[F]^*$ given by
Theorem~\ref{und:four:three}. Then the witnesses must lie in
$F$.\qed

\begin{corollary}
\label{und:four:five}
The existential theory over $F$ is effectively reducible to the
existential consequences of $T[F]^*$ without auxiliary constants, and
to the existential consequences of $T[F]$ without auxiliary functions.
The reduction can be accomplished in linear time.
\end{corollary}

\proof
  From Theorem~\ref{und:one:one} and
  Corollary~\ref{und:four:four}. By
  Theorem~\ref{und:one:three}, the we can use $T[F]$ in place of
  $T[F]^*$.\qed

\begin{corollary}
\label{und:four:six}
If Hilbert's 10th Problem over the rationals is undecidable (as
expected), then the existential consequences of $T[\QQ]$ and
$T[\QQ]^*$, not mentioning auxiliary constants or auxiliary
functions, respectively, are each undecidable. The former can be
reduced to the latter by a linear time reduction.
\end{corollary}

\proof
  Immediate from Corollary~\ref{und:four:five}.\qed


\section{$\forall \forall \forall \exists^*$ consequences
  of $T[F]$}
\label{undecidability:two:section}

We use $\ZZ^+$ for the set of all positive integers, and $\NN$ for the set
of all nonnegative integers.

\begin{lemma}
  \label{und:five:one}
  There exists $\mu,\kappa,\lambda \in R$ such that
\begin{enumerate}
\item For every $n$ in $\NN$, we have $n < \mu$, $\mu^n < \kappa$, and
  $\kappa^n < \lambda$.
\item $[\mu, \infty ) \cap  F = \emptyset$.
\end{enumerate}
\end{lemma}

\proof
By the saturation condition on $F,\RRR$.\qed

We fix $\mu$,$\kappa$,$\lambda$ given by Lemma~\ref{und:five:one}. Let
$K[F,\RRR]$ be the set of all functions $h$ such that
\begin{enumerate}
\item $h \in W[F,\RRR]$.
\item $h$ is the identity on
  $\{\mu,\kappa,\lambda,\mu\kappa,\mu\lambda,\kappa\lambda\}$.
\end{enumerate}
We will build a bijection $h \in K[F,\RRR]$, $h:R \to R$, such that
for all $x \in R$, $1 \le x \le \mu$, the equation
\[
(\kappa+x)(\lambda+x) = \kappa\lambda + \kappa x + \lambda x + x^2
\]
holds in $h^{-1}[\RRR]$ if and only if $x \in \N$. That is, for all $x
\in R$, $1 \le x \le \mu$,
\begin{multline*}
f^{-1}(f(\kappa+x)f(\lambda+x)) = \\
f^{-1}(f(\kappa)f(\lambda)) +
f^{-1}(f(\kappa)f(x)) + f^{-1}(f(\lambda)f(x)) + f^{-1}(f(x)^2) 
\end{multline*}
if and only if $x \in \ZZ^+$. In other words, for all $x \in R$, $1
\le x \le \mu$,
\[
f^{-1}(f(\kappa+x)f(\lambda+x)) =
\kappa\lambda + f^{-1}(\kappa f(x)) +
f^{-1}(\lambda f(x)) + f^{-1}(f(x)^2)
\]
if and only if $x \in \ZZ^+$.

\begin{lemma}
\label{und:five:two}
Let $h \in K[F,\RRR]$, where for every $x$ in $[1,\mu]$, if
\[
h^{-1}(h(\kappa+x)h(\lambda+x)) = \kappa\lambda + h^{-1}(\kappa h(x))
+ h^{-1}(\lambda h(x)) + h^{-1}(h(x)^2),
\]
then $x$ is in $\ZZ^+$. Let $b \not\in \dom(h)$. Then there exists $h
\subseteq_1 h'$ such that $h'(b)$ is defined and for every $x$ in
$[1,\mu]$, if
\[
h'^{-1}(h'(\kappa+x)h'(\lambda+x)) = \kappa\lambda + h'^{-1}(\kappa
h'(x)) + h'^{-1}(\lambda h'(x)) + h'^{-1}(h'(x)^2),
\] 
then $x$ is $\ZZ^+$.
\end{lemma}

\proof
  Let $h$,$b$ be as given. By Lemmas~\ref{und:three:four} and
  \ref{und:two:two}, let $h \subseteq_1 h'$, where $h'(b) \in
  \trans(F,h)$. Note that $\rng(h')\setminus \rng(h)$ consists of the
  nonzero $F$-multiples of $h'(b)$.

  We first show the conclusion for all $ab$, $a \in F \setminus \{ 0
  \}$. We assume
\[
h'^{-1}(h'(\kappa+ab)h'(\lambda+ab)) = \kappa\lambda + h^{-1}(\kappa
h(ab)) + h^{-1}(\lambda h(ab)) + h'^{-1}(h'(ab)^2)
\]
and derive a contradiction.

Clearly $h'^{-1}(h'(ab)^2) = h'^{-1}(a^2h'(b)^2)$ is defined. Since
$h'(b) \in \trans(F,h)$, $a^2h'(b)^2 \in \rng(h')\setminus \rng(h)$.
Hence $a^2h'(b)^2$ is an $F$-multiple of $h'(b)$. This contradicts
that $h'(b) \in \trans(F,h)$.

Finally, we show the conclusion for all $x \in \dom(h)\setminus
\ZZ^+$, $1 \le x \le \mu$. We assume
\begin{equation}
\label{eq:und:three}
h'^{-1}(h'(\kappa+x)h'(\lambda+x)) = \kappa\lambda + h'^{-1}(\kappa
h(x)) + h'^{-1}(\lambda h(x)) + h'^{-1}(h(x)^2)
\end{equation}
and derive a contradiction. By the hypothesis on $h$,
(\ref{eq:und:three}) does not hold with $h'$ replaced by $h$. Hence if
we replace $h'$ by $h$, at least one side of (\ref{eq:und:three}) is
undefined.

First, we claim that $h^{-1}(h(x)^2)$ is defined. Otherwise, $h(x)^2$
is a nonzero $F$-multiple of $h'(b)$. This contradicts that $h'(b) \in
\trans(F,h)$.

Second, we claim that $h^{-1}(\kappa h(x))$ is defined. Otherwise,
$\kappa h(x)$ is a nonzero $F$-multiple of $h'(b)$.

Third, we claim that $h^{-1}(\lambda h(x))$ is defined. Otherwise,
$\lambda h(x)$ is a nonzero $F$-multiple of $h'(b)$.

From these three claims, we see that the right side of
(\ref{eq:und:three}) is defined if we replace $h'$ by $h$. Therefore
$h^{-1}(h(\kappa+x)h(\lambda+x))$ is undefined.

\emph{Case 1.} $h(\kappa+x)$ and $h(\lambda+x)$ are undefined. Then
$h'(\kappa+x)$,$h'(\lambda+x)$ are nonzero $F$-multiples of $h'(b)$.
Since $h'(b) \in \trans(F,h)$, the product $h'(\kappa+x)h'(\lambda+x)
\in \rng(h') \setminus \rng(h)$. Hence $h'(\kappa+x)h'(\lambda+x)$ is
a nonzero $F$-multiple of $h'(b)$. Also $h'(\kappa+x)h'(\lambda+x)$ is
a nonzero $F$-multiple of $h'(b)^2$. This contradicts that $h'(b) \in
\trans(F,h)$.

\emph{Case 2.}  $h(\kappa+x)$ is undefined, but $h(\lambda+x)$ is
defined.  Since $\lambda+x \neq 0$, we have $h(\lambda+x) \neq 0$. Now
$h'(\kappa+x)$ is a nonzero $F$-multiple of $h'(b)$. Since $h'(b) \in
\trans(F,\fld(h))$, $h'(\kappa+x)h(\lambda+x) \in \rng(h')\setminus
\rng(h)$. Hence $h'(\kappa+x)h(\lambda+x)$ is a nonzero $F$-multiple
of $h'(b)$. Therefore $h(\lambda+x) \in F$, and hence $h(\lambda+x) =
\lambda+x \in F$. In particular, $\lambda+x \in F$ and $x \geq 0$.
This contradicts Lemma~\ref{und:five:one}.

\emph{Case 3.}  $h(\kappa+x)$ is defined, $h(\lambda+x)$ is undefined.
Since $\kappa+x \neq 0$, we have $h(\kappa+x) \neq 0$. Now
$h'(\lambda+x)$ is a nonzero $F$-multiple of $h'(b)$. Since $h'(b) \in
\trans(F,\fld(h))$, $h(\kappa+x)h'(\lambda+x) \in \rng(h')\setminus
\rng(h)$. Hence $h(\kappa+x)h'(\lambda+x)$ is a nonzero $F$-multiple
of $h'(b)$. Therefore $h(\kappa+x) \in F$, and hence $h(k+x) =
\kappa+x \in F$. In particular, $\kappa+x \in F$ and $x \geq 0$. This
contradicts Lemma~\ref{und:five:one}.

\emph{Case 4.}  $h(\kappa+x)$ and $h(\lambda+x)$ are defined. Since
$h^{-1}(h(\kappa+x)h(\lambda+x))$ is undefined,
$h(\kappa+x)h(\lambda+x)$ is a nonzero F-multiple of $h'(b)$. This
contradicts that $h'(b) \in \trans(F,h)$.\qed

\begin{lemma}
\label{und:five:three}
Let $h \in K[F,\RRR]$ be such that for every $x$ in $[1,\mu]$, if
\[
h^{-1}(h(\kappa+x)h(\lambda+x)) = \kappa\lambda + h^{-1}(\kappa h(x))
+ h^{-1}(\lambda h(x)) + h^{-1}(h(x)^2),
\]
then $x$ is in $\ZZ^+$. Let
$b \not\in \rng(h)$. Then there exists $h \subseteq_1 h'$ such that
$h'^{-1}(b)$ defined and for every $x$ in $[1,\mu]$, if
\[
h'^{-1}(h'(\kappa+x)h'(\lambda+x)) = \kappa\lambda + h'^{-1}(\kappa
h'(x)) + h'^{-1}(\lambda h'(x)) + h'^{-1}(h'(x)^2),
\] 
then $x$ is in $\ZZ^+$.
\end{lemma}

\proof
  and \ref{und:two:two}, let $h \subseteq_1 h'$, where $h'^{-1}(b) \in
  \trans(F,\fld(h))$. Write $c = h'^{-1}(b)$. Note that $\dom(h')
  \setminus \dom(h)$ consists of the nonzero $F$-multiples of $c$.

  We first show the conclusion for all $ac$, $a \in F \setminus \{ 0
  \}$. We assume
\[
h'^{-1}(h'(\kappa+ac)h'(\lambda+ac)) = \kappa\lambda + h^{-1}(\kappa
h(ac)) + h^{-1}(\lambda h(ac)) + h'^{-1}(h'(ac)^2)
\]
and derive a contradiction. In particular, the assumption implies that
$h'(\kappa+ac)$ is defined, and so $\kappa+ac \in \dom(h)$ or
$\kappa+ac$ is an $F$-multiple of $c$. Both alternatives contradict
that $c \in \trans(F,\fld(h))$.

Finally, we show the conclusion for all $x \in \dom(h) \setminus
\ZZ^+$, $1 \le x \le \mu$. We assume
\begin{equation}
\label{eq:und:four}
h'^{-1}(h'(\kappa+x)h'(\lambda+x)) = \kappa\lambda + h'^{-1}(\kappa
h(x)) + h'^{-1}(\lambda h(x)) + h'^{-1}(h(x)^2)
\end{equation}
and derive a contradiction.

There are five terms in (\ref{eq:und:four}). The four terms other than
$\kappa\lambda$ are each either a nonzero $F$-multiple of $c$ or an
element of $\fld(h)$. Since $c \in \trans(F,\fld(h))$, the ones that
are nonzero $F$-multiples of $c$ must cancel.

We now use the inequalities on $x$,$\mu$,$\kappa$, and $\lambda$. Note
that
\begin{itemize}
\item $ (\kappa+x)(\lambda+x) > \kappa\lambda.$
  \item $h'((\kappa+x)(\lambda+x)) > h'(\kappa\lambda) = \kappa\lambda. $
  \item $ h'^{-1}(h'(\kappa+x)h'(\lambda+x)) > h'^{-1}(\kappa\lambda) = \kappa\lambda. $
  \item $ x \le \mu.$
  \item $ h(x) \le h(\mu) = \mu.$
  \item $ \kappa h(x) \le \mu\kappa.$
  \item $ h'^{-1}(\kappa h(x)) \le h'^{-1}(\mu\kappa) = \mu\kappa.$
  \item $ \lambda h(x) \le \mu\lambda.$
  \item $ h'^{-1}(\lambda h(x)) \le h'^{-1}(\mu\lambda) = \mu\lambda.$
  \item $ h(x)^2 \le \mu^2 < \kappa.$
  \item $ h'^{-1}(h(x)^2) \le h'^{-1}(\kappa) < \kappa.$
  \item $ h'^{-1}(h'(\kappa+x)h'(\lambda+x)) > \kappa\lambda > \mu\kappa +
  \mu\lambda + \kappa \geq h'^{-1}(\kappa h(x)) + h'^{-1}(\lambda h(x)) + h'^{-1}(h(x)^2)$.
\end{itemize}
It is now obvious that the terms that are nonzero $F$-multiples of $c$
cannot include $h'^{-1}(h'(\kappa+x)h'(\lambda+x))$.

This leaves $h'^{-1}(\kappa h(x))$, $h'^{-1}(\lambda h(x))$,
$h'^{-1}(h(x)^2)$ as the terms that might be nonzero $F$-multiples of
$c$. Using the above, we have
\begin{itemize}
\item $h'^{-1}(h(x)^2) < \kappa. $
\item $h'^{-1}(\kappa h(x)) \le \mu\kappa.$
  \item $ h'^{-1}(\lambda h(x)) \le \mu\lambda.$ 
  \item $ x \geq 1.$
  \item $ h(x) \geq h(1) = 1.$
  \item $ \kappa h(x) \geq \kappa.$
  \item $ h'^{-1}(\kappa h(x)) \geq h'^{-1}(\kappa) = \kappa.$ 
  \item $ h(x) \geq h(1) = 1.$
  \item $ \lambda h(x) \geq \lambda.$
  \item $ h'^{-1}(\lambda h(x)) \geq h'^{-1}(\lambda) = \lambda.$ 
\end{itemize}
Hence
\begin{itemize}
  \item $ h'^{-1}(h(x)^2) < \kappa.$
  \item $ \kappa \le h'^{-1}(\kappa h(x)) \le \mu\kappa.$
  \item $ \lambda \le h'^{-1}(\lambda h(x)).$
\end{itemize}
It is now clear that none of $h'^{-1}(\kappa h(x))$, $h'^{-1}(\lambda
h(x))$, $h'^{-1}(h(x)^2)$ can be a nonzero $F$-multiple of $c$. Hence
\[
h'^{-1}(h'(\kappa+x)h'(\lambda+x)), \quad h'^{-1}(\kappa h(x)), \quad
\mbox{and} \quad h'^{-1}(\lambda h(x)), h'^{-1}(h(x)^2)
\]
all lie in $\dom(h)$. Therefore
\[
h'(\kappa+x)h'(\lambda+x), \quad \kappa h(x), \quad \lambda h(x), 
\quad \mbox{and} \quad h(x)^2
\] 
lie in $\rng(h)$. We claim that $h'(\kappa+x), h'(\lambda+x) \in
\rng(h)$. To see this, first suppose both are not in $\rng(h)$. Then
$\kappa+x$ and $\lambda+x$ are $F$-multiples of $c$, and so
$(\kappa+x)(\lambda+x)$ is of the form $aa'c^2$, where $a,a' \in F$.
This contradicts the fact that $c$ is in $\trans(F,h)$.

Now suppose one of them, say, by symmetry, $h'(\kappa+x)$, is an
$F$-multiple of $c$, and the other, $h'(\lambda+x)$, lies in
$\rng(h)$. Since $\lambda+x \neq 0$, we have $h'(\lambda+x) \neq 0$.
Then $h'(\kappa+x)h'(\lambda+x)$ is of the form $acu$, where $a \in F
\setminus \{ 0 \}$ and $u \in \rng(h)$. But $h'(\kappa+x)h'(\lambda+x)
\in \rng(h)$. Hence $acu \in \rng(h)\setminus \{0\}$. This contradicts
the fact that that $c$ is in $\trans(F,h)$.

From $h'(\kappa+x), h'(\lambda+x) \in \rng(h)$, we obtain that
$\kappa+x, \lambda+x \in \dom(h)$. Thus we see that both sides of
(\ref{eq:und:four}) are defined if we replace $h'$ by $h$. Hence
(\ref{eq:und:four}) holds with $h'$ replaced by $h$. This is a
contradiction.\qed

We want to iterate Lemmas~\ref{und:five:two} and \ref{und:five:three},
but we first need to deal with the base case. Let
\begin{multline*}
S = \{\kappa+x: x \in \ZZ^+\} \cup \{\lambda+x: x \in \ZZ^+\}
\cup \{\kappa\lambda+\kappa x+\lambda x+x^2 \st x \in \ZZ^+\} \cup\\
\{1,\mu,\kappa,\lambda,\mu\kappa,\mu\lambda,\kappa\lambda\}.
\end{multline*}
Let $S'$ be the set of all $F$-multiples of elements of $S$.

\begin{lemma}
\label{und:five:four}
Let $x \in S'$, $1 \le x \le \mu$. If $\kappa+x \in S'$ then $x \in
\ZZ^+$. If $\lambda+x \in S'$ then $x \in \ZZ^+$.
\end{lemma}

\proof
  Let $x$ be as given. Suppose $\kappa+x \in S'$. Since $\kappa+x <
  2\kappa$, clearly $\kappa+x$ is not a nonzero $F$-multiple of any
  element of
\[
\{\lambda+x \st x \in \ZZ^+\} \cup
\{\kappa\lambda+\kappa x+\lambda x+x^2 \st x \in \ZZ^+\} \cup
\{\lambda,\mu\kappa,\mu\lambda,\kappa\lambda\}.
\]
Since $\kappa+x$ is greater than every $\mu^n$, $n \in \ZZ^+$,
$\kappa+x$ is not a nonzero $F$-multiple of any element of
$\{1,\mu\}$.

Now suppose $\kappa+x$ is an $F$-multiple of $\kappa+y$, $y \in \NN$.
Write $\kappa+x = a(\kappa+y)$, $a \in F$. Then $\kappa =
(ay-x)/(1-a)$ or $a = 1$. Now $|ay-x| \le |ay| + |x| \le \mu + \mu =
2\mu$. Also $1/|1-a| \le \mu$ or $a = 1$. Hence $\kappa \le 2\mu^2$ or
$a = 1$. Therefore $a = 1$. Hence $\kappa+x = \kappa+y$, and $x = y$.
Therefore $x \in \ZZ^+$.

Suppose $\lambda+x \in S'$. Since $\lambda+x < 2\lambda$, clearly
$\lambda+x$ is not a nonzero $F$-multiple of any element of
$\{\mu\lambda,\kappa\lambda\} \cup
\{\kappa\lambda+\kappa x+\lambda x+x^2 \st x \in \NN\}$. Since
$\lambda+x$ is greater than every $\kappa^n$, $n \in \ZZ^+$,
$\lambda+x$ is not a nonzero $F$-multiple of any element of
$\{\kappa+x: x \in \ZZ^+\} \cup \{1,\mu,\kappa,\mu\kappa\}$.

Now suppose $\lambda+x$ is a nonzero $F$-multiple of $\lambda+y$, $y
\in \ZZ^+$. Argue as above that $x \in \ZZ^+$.\qed

\begin{lemma}
  \label{und:five:five} There exists a bijection $h \in K[F,\RRR]$,
  $h:R \to R$, such that the following holds. For all $x \in \dom(h)$
  with $1 \le x \le \mu$, we have 
\[
h^{-1}(h(\kappa+x)h(\lambda+x)) =
  \kappa\lambda + h^{-1}(\kappa h(x)) + h^{-1}(\lambda h(x)) +
  h^{-1}(h(x)^2)
\]
 if and only if $x$ is in $\ZZ^+$.
\end{lemma}

\proof
  Let $h$ be the identity function on $S'$. Obviously $h \in
  K[F,\RRR]$. By Lemma~\ref{und:five:four}, for all $x \in \dom(h)$
  such that $1 \le x \le \mu$, if
\[
h^{-1}(h(\kappa+x)h(\lambda+x)) = \kappa\lambda + h^{-1}(\kappa h(x))
+ h^{-1}(\lambda h(x)) + h^{-1}(h(x)^2)\] then $x \in \ZZ^+$. This is
because for the relevant $x$, if $h(\kappa+x)$ is defined then $x \in
\ZZ^+$.

For the reverse, let $x \in \ZZ^+$, and note that
\begin{multline*}
h^{-1}(h(\kappa+x)h(\lambda+x)) = h^{-1}((\kappa+x)(\lambda+x)) =
h^{-1}(\kappa\lambda+\kappa x+\lambda x+x^2) =\\
\kappa\lambda+\kappa x+\lambda x+x^2.
\end{multline*}
So
\begin{multline*}
\kappa\lambda + h^{-1}(\kappa h(x)) + h^{-1}(\lambda h(x)) +
h^{-1}(h(x)^2) = \\
\kappa\lambda + h^{-1}(\kappa x) + h^{-1}(\lambda x) +
h^{-1}(x^2) = \kappa\lambda+\kappa x+\lambda x+x^2.
\end{multline*}\qed

\begin{lemma}
\label{und:five:six}
There exists a bijection $h \in K[F,\RRR]$, $h:R \to R$, such that the
following holds. For all $x \in R$ with $1 \le x \le \mu$, we have
\[
h^{-1}(h(\kappa+x)h(\lambda+x)) = \kappa\lambda + h^{-1}(\kappa h(x))
+ h^{-1}(\lambda h(x)) + h^{-1}(h(x)^2)
\]
if and only if $x$ is in $\ZZ^+$.
\end{lemma}

\proof
  Start with the $h$ given by Lemma~\ref{und:five:five}, and iterate
  Lemmas~\ref{und:five:two} and \ref{und:five:three}, diagonalizing
  over the countably many elements of $R$.\qed

\begin{theorem}
\label{und:five:seven}
There is a model $\mdl M$ of $T[F]^*$ with domain $R$, with the same
$0,1,+,<$ as $\RRR$, with three elements $\mu,\kappa,\lambda$ such
that the following holds. For all $x \in R$ with $1 \le x \le \mu$, we
have $(\kappa+x)(\lambda+x) = \kappa\lambda+\kappa x+\lambda x+x^2$ if
and only if $x$ is in $\ZZ^+$. In this equation, we use the
multiplication of $\mdl M$.
\end{theorem}

\proof
  By Theorem~\ref{und:two:one} and Lemma~\ref{und:five:six}.\qed

We say that a quadruple $\la M,\mu,\kappa,\lambda \ra$ has property
(*) if and only if
\begin{enumerate}
\item $\mdl M$ is a model of $T[F]^*$.
\item $\mu,\kappa,\lambda \in \dom(\mdl M)$.
\item The $x \in \dom(M)$ for which $1 \le x \le \mu$ and
  $(\kappa+x)(\lambda+x) = \kappa\lambda+\kappa x+\lambda x+x^2$ contain
  $1$ and are closed under $+1$.
\end{enumerate}
  There is the stronger property (**)
  of $\la \mdl M,\mu,\kappa,\lambda \ra$ that asserts the following.
\begin{enumerate}
\item $\mdl M$ is a model of $T[F]^*$.
\item $\mu,\kappa,\lambda \in \dom(\mdl M)$.
\item The $x \in \dom(M)$ for which $1 \le x \le \mu$ and
  $(\kappa+x)(\lambda+x) = \kappa\lambda+\kappa x+\lambda x+x^2$ are
  exactly the positive integers in $\mdl M$.
\end{enumerate}

\begin{corollary}
  \label{und:five:eight}
  Let $D$ be a Diophantine equation over the positive integers. Then
  $D$ has a solution in nonnegative integers if and only if the
  following holds. For all quadruples $\la \mdl M,\mu,\kappa,\lambda
  \ra$ with property (*), $D$ has a solution over the $x$ such that $1
  \le x \le \mu$ and $(\kappa+x)(\lambda+x) =
  \kappa\lambda+\kappa x+\lambda x+x^2$.
\end{corollary}

\proof
  Let $D$ be as given. Suppose $D$ has a solution in the positive
  integers. Let $\la \mdl M,\mu,\kappa,\lambda \ra$ have property (*).
  Then the $x$ such that $1 \le x \le \mu$ and $(\kappa+x)(\lambda+x)
  = \kappa\lambda+\kappa x+\lambda x+x^2$ must contain the positive
  integers.

  Conversely, suppose that for all quadruples $\la
  M,\mu,\kappa,\lambda \ra$ with property (*), $D$ has a solution over
  the $x$ such that $1 \le x \le \mu$ and $(\kappa+x)(\lambda+x) =
  \kappa\lambda+\kappa x+\lambda x+x^2$. By
  Theorem~\ref{und:five:eight}, there exists $\la \mdl
  M,\mu,\kappa,\lambda\ra$ with property (**). Hence $D$ has a
  solution over the positive integers.\qed

\begin{theorem}
\label{und:five:nine}
The set of consequences of $T[F]^*$ without auxiliary constants, and
of $T[F]$ without auxiliary functions, is undecidable. In fact, the
set of $\forall \forall \forall \exists^*$ consequences of
$T[F]^*$ without auxiliary constants, and of $T[F]$ without auxiliary
functions, is complete r.e.
\end{theorem}

\proof
  We use Corollary~\ref{und:five:eight} and that Hilbert's 10th
  problem over $\ZZ^+ $is complete r.e. We can express
\begin{quote}
  $\la \mdl M,\mu,\kappa,\lambda \ra$ has property (*)
\end{quote}
as the formula $\ph(\mu,\kappa,\lambda)$ given by
\begin{multline*}
(\kappa+1)(\lambda+1) = \kappa\lambda+\kappa+\lambda+1 \mathop{\land} \\
  \fa x((1 \le x \le \mu \land  (\kappa+x(\lambda+x) =
  \kappa\lambda+\kappa x+\lambda x+x^2) \limplies \\
(1 \le x+1 \le \mu \land (\kappa+(x+1))(\lambda+(x+1)) = \kappa\lambda+\kappa(x+1)+\lambda(x+1)+(x+1)^2)).
\end{multline*}
Then we can write 
\begin{quote}
  for all quadruples $\la \mdl M,\mu,\kappa,\lambda \ra$ with property
  (*), $D$ has a solution over the $x$ such that $1 \le x \le \mu$ and
  $(\kappa+x)(\lambda+x) = \kappa\lambda+\kappa x+\lambda x+x^2$
\end{quote}
as the assertion that
\begin{multline*}
\fa{\mu,\kappa,\lambda} (\ph(\mu,\kappa,\lambda) \limplies \\
\mbox{$D$ has
  a solution over the $x$ such that $1 \le x \le \mu$ and}\\
\mbox{
  $(\kappa+x)(\lambda+x) = \kappa\lambda+\kappa x+\lambda x+x^2$)}
\end{multline*}
is provable in $T[F]^*$.  Note that the sentence above is in the form
$\forall \forall \forall \exists^*$. By Theorem~\ref{und:one:three},
we can replace $T[F]$ by $T[F]^*$.\qed


\section{Avoiding disjunctions}
\label{avoiding:section}

In Section~\ref{decidability:section}, we saw that the universal
fragment of $T[\QQ]$ is decidable. The proof, however, involves a
complex reduction to the language of real closed fields. As a result,
the procedure is of little practical importance: $T[\QQ]$ is weaker
than the theory of real closed fields, our decision procedure works
for only the universal fragment of the language, and it does so less
efficiently than procedures for the corresponding fragment of real
closed fields. The procedure we describe is in no sense more
extensible to larger languages than procedures for real closed
fields. It may therefore seem as though we have taken a step in the
wrong direction.

We maintain, however, that the analysis provides guidance in
designing heuristic procedures for the reals that address the aims
outlined in Section~\ref{introduction:section}. An obvious strategy
for capturing inferences like the ones described there is to work
backwards from the desired conclusion, using the obvious monotonicity
laws. For example, when the terms $s$, $t$, and $u$ are known to be
positive, one can prove $s t \leq u v$ by proving $s \leq u$ and $t
\leq v$. The examples presented in Section~\ref{introduction:section}
can be verified by iteratively applying such rules.

There are drawbacks to such an approach, however. For one thing,
excessive case splits can lead to exponential blowup; e.g.~one can
show $s t > 0$ by showing that $s$ and $t$ are either both strictly
positive or both strictly negative. And the relevant monotonicity
inferences are generally nondeterministic: one can show $r + s + t >
0$ by showing that two of the terms are nonnegative and the third is
strictly positive, and one can show $r + s < t + u + v + w$, say, by
showing $r < u$, $s \leq t + v$, and $0 \leq w$.

In ``straightforward'' inferences that arise in practice, however,
sign information is typically available. This is the case with the
examples in Section~\ref{introduction:section}, where all the relevant
terms are easily seen to be positive. It is also the case with the
following representative example, taken from the first author's
formalization of the prime number theorem~\cite{avigad:et:al:unp}:
verify
\[
(1 + \frac{\varepsilon}{3(C + 3)}) \cdot n < K x
\]
using the hypotheses
\[
\begin{split}
& n \leq (K / 2) x \\
& 0 < C \\
& 0 < \varepsilon < 1.
\end{split}
\]
This is easily verified by noting that $1 + \frac{\varepsilon}{3(C +
  3)}$ is strictly less than $2$, and so the product with $n$ is
strictly less than $2 (K / 2) x = K x$. In this case, backchaining
does not work, unless one thinks of replacing $K x$ by $2((K / 2)x)$ in
the goal inequality.

This example suggests that some form of forward search may be more
fruitful: starting from the hypotheses, iteratively derive useful
consequences, until the goal is obtained. Alternatively, we negate the
conclusion and add it to the list of hypotheses, and then iteratively
derive consequences until we obtain a contradiction. Our analysis
shows that if we separate terms, we can in fact use $\Tadd[F]$ and
$\Tmult[F]$ independently to derive consequences, and that we only
have to consider consequences in the language of $\Tcommon[F]$. This
procedure is complete for the universal consequences of $T[F]$, and
works equally well if we combine other local decision procedures for
languages that are disjoint except for $=$ and $\leq$.

But what consequences shall we look for? Once again, our analysis
shows us that a single well-chosen interpolant suffices: if we pick
the right $\theta$, $\Tadd[F]$ will be able to derive $\theta$ from
our initial set of hypotheses, while $\Tmult[F]$ will be able to prove
$\lnot \theta$. According to Proposition~\ref{interpolant:prop} and
the discussion after it, we can assume, without loss of generality,
that $\theta$ is a conjunction of disjunctions of literals of the form
$x_i < a x_j, x_i \leq a x_j, x_i > a x_j, x_i \geq a x_j$, and
comparisons between variables and constants in $F$. As a result, if
the initial sequence of hypotheses can be refuted, there is a sequence
$\theta_1, \theta_2, \ldots, \theta_n$ of disjunctions of atomic
formulas of the form above, such that $\Tadd[F]$ proves each formula
$\theta_i$ from the initial set of hypotheses, and $\Tmult[F]$ proves a
contradiction from these hypotheses and $\theta_1, \ldots, \theta_n$.
Of course, the situation is symmetric, so we can just as well switch
$\Tadd[F]$ and $\Tmult[F]$ in the previous assertion.

This reduces the task to that of deriving appropriate disjunctions
$\theta_i$ of atomic formulas $x_i \leq a x_j$ from the initial
hypotheses. The problem is that there are always infinitely many
disjunctions that one can prove, and it may not be clear which ones
are likely to be useful. For example, from $x + y \geq 0$, $\Tadd[F]$ can
prove $x \geq a \lor y \geq -a$ for any $a$, and, a priori, any of
these may be useful to $\Tmult[F]$.

One solution is simply to ignore disjunctions. By
Proposition~\ref{separating:variables:prop}, with some initial case
splits we can reduce the problem of proving a universal formula to
refuting a finite number of sets of formulas of the form $\Deltaadd
\cup \Deltamult \cup \Deltacommon$, where
\begin{itemize}
\item $\Deltaadd$ is a set of formulas of the form $x_i = t$, where
  $t$ is a term in the language of $\Tadd[F]$;
\item $\Deltamult$ is a set of formulas of the form $x_i = t$, where
  $t$ is a term in the language of $\Tmult[F]$;
\item $\Deltacommon$ is a set of formulas of the form $x_i < a x_j,
  x_i \leq a x_j, x_i > a x_j, x_i \geq a x_j$, or a comparison
  between a variable and a constant.
\end{itemize}

\begin{definition}
  Let $\Delta = \Deltaadd \cup \Deltamult \cup \Deltacommon$ be as
  above. Say $T[F]$ \emph{refutes $\Delta$ without case splits} if
  there is a sequence of atomic formulas $\theta_0, \ldots,
  \theta_{2n}$ such that the following hold:
\begin{itemize}
\item for $m < 2n$, $\theta_m$ has the same form as the formulas in
  $\Deltacommon$; 
\item $\theta_{2n}$ is $\bot$;
\item for each $m < n$, 
\[
\Tadd[F] \cup \Deltaadd \cup \Deltacommon \cup \{ \theta_0, \ldots,
    \theta_{2m-1} \} \proves \theta_{2m};
\]
\item for each $m < n$, 
\[
\Tmult[F] \cup \Deltamult \cup \Deltacommon \cup \{ \theta_0, \ldots,
    \theta_{2m} \} \proves \theta_{2m+1}.
\]
\end{itemize}
\end{definition}
In other words, $T[F]$ refutes $\Delta$ without case splits if
$\Tadd[F]$ and $\Tmult[F]$ can iteratively augment a database of
derivable atomic formulas in the common language until a contradiction
is reached. This is a proper restriction on the theories $T[F]$, which
is to say, there are sets $\Delta$ that can be refuted by $T[F]$, but
not without case splits. It takes some effort, though, to cook up an
example. Here is one. Let
\[
\Deltaadd = \{ x + y \geq 2, w + z \geq 2 \}
\]
From this, $\Tadd[F]$ proves $(x \geq 1 \lor y \geq 1) \land (w \geq 1
\lor z \geq 1)$. Let
\[
\Deltamult = \{ u x^2 < u x, u y^2 < uy, u w^2 > u w, u z^2 > u z \}.
\]
From this, $\Tmult[F]$ proves $u > 0 \lor u < 0$, and hence $(x < 1 \land
y < 1) \lor (w < 1 \land z < 1)$. As a result, $T[F]$ refutes
$\Deltaadd \cup \Deltamult$. But one can check that there are
no atomic consequences involving the common variables, $x, y, z$ and
$w$, that follow from either set. (Strictly speaking, our
characterization of $\Delta$ has us using new variables to name
the additive and multiplicative terms in $\Deltaadd$ and $\Deltamult$,
respectively, and then putting the comparisons in $\Deltacommon$. But
the net effect is the same.)

Situations like this are contrived, however, and we expect that
focusing on atomic consequences will be effective in many ordinary
situations. The following proposition provides some encouragement.

\begin{proposition}
\label{best:add:prop}
Let $\Delta$ be a set of atomic formulas in the language of
$\Tadd[F]$. Let $u$ and $v$ be any two variables. Then there is a
consequence, $\theta$, of $\Tadd[F] \cup \Delta$ in the language of
$\Tcommon[F]$, involving only $u$ and $v$, that implies all the
consequences of the form $u < a v$, $u \leq a v$, $v < au$, or $v \leq
au$ that can be derived from $\Tadd[F] \cup \Delta$. In fact, $\theta$
can be expressed as a conjunction of at most two formulas of the form
$u < a v$, $u \leq a v$, $u > av$, $u \geq av$, $v < 0$, $v \leq 0$,
$v > 0$, or $v \geq 0$.
\end{proposition}

\proof
  Use a linear elimination procedure to eliminate all variables except
  for $u$ and $v$ from $\Delta$. The result is a set of linear
  inequalities involving $u$ and $v$, which implies every other
  relation between $u$ and $v$ that is derivable from $\Tadd[F] \cup
  \Delta$. (If a relation is not a consequence of the resulting set of
  linear inequalities, its negation is consistent with them, and hence
  with $\Tadd[F] \cup \Delta$.) This set of linear inequalities
  determines a convex subset of the cartesian plane. Considering
  extremal points, one can determine the minimal intersection of at
  most two half planes through the origin that includes this convex
  subset.\qed

An efficient algorithm for determining the convex polygon determined
by a sequence of half-planes can be found in \cite[Section
4.2]{de:berg:et:al:00}. Keep in mind that there may be \emph{no}
nontrivial consequences of $\Delta$, in which case we can take
$\theta$ to be the empty conjunction, $\top$. Or $\Delta$ may
contradictory, in which case we can take $\theta$ to be $\bot$, or $v
< 0 \land v > 0$. Furthermore, $\theta$ may not be strong enough to
determine whether $u$ and $v$ are positive, negative, etc. In that
case, as in the discussion after Proposition~\ref{interpolant:prop},
determining whether one inequality is stronger than another can be
confusing. For example, $\theta$ may be $u > 2v \land u > 3v$; in the
absence of sign information, neither conjunct is stronger. If one adds
the information $v > 0$, $\theta$ becomes $v > 0 \land u > 3 v$.

On the multiplicative side, we have to assume we know the signs of the
variables, and that $F$ is closed under $n$th roots. 

\begin{proposition}
\label{best:mult:prop}
Let $\Delta$ be a set of atomic formulas in the language of
$\Tmult[F]$. Assume that for each variable $x$ occurring in $\Delta$,
$\Delta$ contains either the formula $x > 0$ or the formula $x < 0$.
Assume also that $F$ is closed under $n$th roots of positive numbers
for positive integers $n$.  Let $u$ and $v$ be any two variables. Then
there is a consequence, $\theta$, of $\Tmult[F] \cup \Delta$ in the
language of $\Tcommon[F]$, involving only $u$ and $v$, that implies
all the consequences of the form $u < a v$, $u \leq a v$, $v < au$, or
$v \leq au$ that can be derived from $\Tmult[F] \cup \Delta$. In fact,
$\theta$ can be expressed as a conjunction of at most two formulas of
the form $u < a v$, $u \leq a v$, $u > av$, or $u \geq av$.
\end{proposition}

The good news is that the proof is even easier in this case.

\proof
Introduce a new variable $w$, and the equation $w = u/v$. Eliminate
all variables except for $w$. The result is a set of inequalities of
the form $w < a$, $w \leq a$, $w > a$, and $w \geq a$, of which we can
choose the strongest and then replace $w$ by $u / v$.\qed

The requirement that we have sign information on the variables is
generally needed to carry out the elimination procedure for
$\Tmult[F]$. We can always ensure that this information is present
using case splits, though this can be computationally expensive. The
requirement that $F$ is closed under taking roots is also needed for
the conclusion; for example, from $\{ u > 0, u^2 > 2 v^2 \}$ we would
like to conclude $u > \sqrt{2} v$. For practical purposes, however, we
will suggest, in the next section, that one should choose $\QQ$ for
$F$ in an implementation, and avoid case splits. In that case, we can
only hope for an approximation to Proposition~\ref{best:mult:prop}.
For example, when trying to put a multiplicative equation in pivot
form, if we do not have sufficient sign information to determine the
appropriate direction of an inequality, we can simply ignore this
equation. And when required to take $n$th roots at the very end of the
procedure, we can rely on crude approximations, such as $\sqrt[n]{a} >
1$ whenever $a > 1$. Once again, we expect that even with these
concessions, the resulting procedure will be helpful in verifying
commonplace inferences.

This strategy, then, will form the basis for the heuristic procedure
that we will suggest in the next section. We leave open one
interesting theoretical question, though: is it decidable whether a
theory $T[F]$ can refute a set $\Delta$ without case splits? The proof
of Theorem~~\ref{upper:bound:one:thm} shows that trying to refute the
set $\Delta$ corresponding to $x^2 + 2x - 1 < 0$ leads to an infinite
iteration, so the obvious search procedure is not guaranteed to
terminate.


\section{Towards a heuristic procedure}
\label{heuristic:section}

In this section, we discuss some possible avenues towards developing
heuristic decision procedures, based on the analysis we have provided
here. We are, of course, sensitive to the tremendous gap between neat
decidability results and heuristic procedures that work well in
practice. But we expect that the former can serve as a useful guide in
the development of the latter, by clarifying the inherent
possibilities and limitations of the method, and separating heuristic
issues from theoretical ones. Of course, different heuristic
approaches will have distinct advantages and disadvantages, and so
different procedures can be expected to work better in different
domains. We expect the type of algorithm we propose here to be
fruitful for the kinds of examples discussed in
Section~\ref{introduction:section}.

Given a quantifier-free sequent in the language of $T[\QQ]$, first,
put all terms in normal form, as described in
Section~\ref{normal:forms:section}. This will make it possible to
identify subterms that are provably equal. For that purpose, one can
use the simpler normal form described at the end of
Section~\ref{normal:forms:section}.

Next, use new variables, recursively, to name additive and
multiplicative subterms. These will form the sets $\Deltaadd$ and
$\Deltamult$. With these renamings, the original sequent will be
equivalent to one in the language of $\Tcommon[\QQ]$.

Convert the resulting sequent to a finite sequence of sets
$\Deltacommon$ of inequalities $x < ay$, $x \leq ay$, $x > ay$, $x \ge
ay$, to be refuted. For example, proving the sequent
\[
x = y, w < z \Rightarrow u < v
\] 
amounts to refuting the set $\Deltacommon$ of formulas
\[
\{ x \geq y, x \leq y, w < z, v \leq u \}.
\]
Note that the equality in the hypothesis is replaced by two
inequalities. This seems to be a reasonable move, since with
$\Deltaadd$ and $\Deltamult$, $x$ and $y$ may name complex terms; we
imagine that this procedure will be called after obvious
simplifications and rewriting have been performed. Also note that the
task of proving an equality $u = v$ splits into two tasks, namely,
refuting $u > v$ and refuting $u < v$. Again, this seems reasonable,
since we envision this procedure being called when direct methods for
proving equalities have failed. 

Now, try to refute each set $\Deltacommon$, with the following
iterative procedure. First, for each pair of variables $x, y$ in
$\Deltacommon$, use $\Tadd[\QQ] \cup \Deltacommon$ to derive new or
stronger inequalities of the form $x < ay$, $x \leq ay$, $x > ay$, or
$x \geq ay$, as well comparisons between $x$ and constants for each
variable $x$. Add the new inequalities to $\Deltacommon$, removing
ones that are subsumed by the new information. $\Deltacommon$ can be
represented as a table of comparisons for each pair $\{ x, y \}$ (for
each pair, at most two formulas need to be stored), as well as a table
of comparisons with constants for each variable $x$. Even though the
procedure implicit in Proposition~\ref{best:add:prop} invokes a
linear elimination procedure (see the discussion and references in
Section~\ref{fragments:section}), the work can be shared when cycling
through all possible pairs. For example, to determine all inequalities
obtainable from a set with $n$ variables, eliminate the first
variable, $x$, and recursively determine all the inequalities
obtainable from the resulting set with $n$ variables; then determine
all the inequalities that can be obtained with $x$ and one other
variable.  Furthermore, at least initially, for most pairs no
information will be available at all, and so will be eliminated
quickly.  We expect that for the types of problems that arise in
ordinary practice, the number of variables and named subterms will be
small enough to make the procedure manageable. If not, heuristics can
be used to focus attention on pairs that are likely to provide useful
information.

Do the same with $\Tmult[\QQ] \cup \Deltamult$. First, use the
information in $\Deltamult$ to determine the variables for which one
has comparisons with $0$. For a defining equation such as $u = x^2
y^4$, the multiplicative procedure can infer $u \geq 0$ at the start,
and add it to $\Deltacommon$ for possible use by the additive
procedure. With limited sign information on the variables, let the
procedure for $\Tmult[\QQ] \cup \Deltamult$ do the best it can to
eliminate variables. If it cannot make use of an inequality $x^k s <
t$ to eliminate $x$ because the sign of $s$ is not known, simply
ignore the inequality at this stage. It may become useful later on, if
the sign of $s$ becomes known.

Iterate the additive and multiplicative steps, until one of $\Deltaadd
\cup \Deltacommon$ or $\Deltamult \cup \Deltacommon$ yields a
contradiction. Of course, there is the question as to when to give up.
One can certainly report failure when no new inequalities have been
derived. But as noted at the end of Section~\ref{avoiding:section},
nonterminating iterations are possible; in that case, the procedure
can simply give up after a fixed amount of time, or rely on the user
to halt the procedure.


\section{Extending the heuristic}
\label{extensions:section}

There are many ways that one may extend the proposal in the previous
section. These fall into general classes.

\medskip

\noindent \emph{Improvements to the heuristic.} There are likely to be
better ways of searching for useful comparisons between terms. For
example, one can have a list of ``focus'' formulas -- initially, one
wants to include the goal formula as a focus formula -- and search for
inequalities between subterms of those. Also, one does not need to
search for comparisons between two variables unless information has
been added to $\Deltacommon$ since the last such search that could
potentially yield new information. Thus, a wise choice of data
structures and representations of information in the database may
yield significant improvements.

\medskip

\noindent \emph{Extensions to stronger fragments of $T[\QQ]$.} The
procedure we have described does not try to derive disjunctions, which
requires potentially costly case splits. Are there situations in which it
makes sense to introduce such splits? For example, it may be useful to
split on the sign of a variable, $x \geq 0 \lor x < 0$; or to split on
a comparison between two variables, $x \geq y \lor x < y$, where $x$
and $y$ name terms in the search.

\medskip 

\noindent \emph{Conservative extensions of $T[\QQ]$.} The functions
which return $n$th roots, absolute value, minimums, and maximums can
all be defined in $T[\QQ]$, and it would be useful to extend the
heuristic to languages that include these. But, as discussed at
the end of Section~\ref{normal:forms:section}, one has to either
introduce case splits at the outset to simplify terms appropriately,
or simplify a term like $\sqrt{x^2}$ to $x$ when $x \geq 0$ is
determined in the course of the search. What is the best way to handle
such extensions?

\medskip

\noindent \emph{Nonconservative extensions of $T[\QQ]$, in the same
  language.} An obvious shortcoming of $T[\QQ]$ is that it fails to
capture straightforward inferences that are easily obtained using
distributivity. On the other hand, using distributivity to simplify an
expression before calling a decision procedure for $T[\QQ]$ can erase
valuable information; for example, after simplification, $T[\QQ]$ can
no longer verify $(x + 1)^2 \geq 0$. A better strategy is to perform
such simplifications as the search proceeds, when occasion seems to
warrant it, perhaps retaining the factored versions as well.

As noted in Section~\ref{introduction:section}, it is reasonable to
claim that any validity that requires complex factoring falls outside
the range of the ``obvious,'' and hence outside the scope of the
problem we are concerned with here.  But one would expect a good
procedure to multiply through in at least some contexts, i.e.~only use
distributivity in the ``left-to-right'' direction to simplify
expressions at hand. The question is how to work this in to the
procedures described below in a principled way. It would also be nice
to have a better theoretical framework to discuss provability with
equalities ``applied only in the left-to-right direction.''

\medskip

\noindent \emph{Amalgamating other decision and heuristic procedures.}
A major advantage of the method described in
Section~\ref{heuristic:section} is that it can easily be scaled to
allow other procedures to add facts to the common database. For
example, one can easily make use of the equivalence $x < y \liff f(x)
< f(y)$ for a strictly monotone function $f$. One can similarly add
procedures that make use of straightforward properties of
transcendental functions like $\exp$, $\ln$, $\fn{sin}$, $\fn{cos}$,
and so on.

\medskip

\noindent \emph{Extending the overlap.} Just as one might make use of
limited forms of distributivity, one can add restricted uses of laws
like $e^{x + y} = e^x e^y$, for the exponential function.

\medskip

\noindent \emph{Handling subdomains, like $\ZZ$ and $\QQ$, and
  extended domains, like $\CC$.} For example, it is known that the
linear theory of the reals with a predicate for the integers is
decidable (see, for example, \cite{weispfenning:99b}). Handling mixed
domains involving $\NN$, $\ZZ$, $\QQ$, $\RR$, and/or $\CC$ is an
important challenge for heuristic procedures.

\medskip


\section{Conclusions}
\label{conclusions:section}

In order to obtain useful methods for verifying inferences in
nontrivial mathematical situations, undecidability and infeasibility
should encourage one to search for novel ways of delimiting
manageable, restricted classes of inferences that include the ones
that come up in ordinary mathematical practice. We hope our study of
inferences involving inequalities between real-valued expressions that
can be verified without using distributivity is an interesting and
fruitful investigation along these lines. We also feel that the
paradigm of amalgamating decision or heuristic procedures when there
is nontrivial overlap between the theories is an important one for
automated reasoning.

However, we expect that similar investigations can be carried out in
almost any mathematical domain. This yields both theoretical and
practical challenges. On the theoretical side, for example, there are
questions of decidability and complexity. On the practical side, there
is always the question of how to implement proof searches that work
well in practice. As a result, we feel that this type of research
represents a promising interaction between theory and practice.


\bibliographystyle{plain} 
\bibliography{proofthry,automated}

\end{document}